\documentclass{emulateapj}
\usepackage{epsfig}
\usepackage{natbib}
\usepackage{rotating}
\newcommand{\degree}{\mbox{$^{\circ}$}}
\newcommand{\am}{\mbox{$^{\prime}$}}
\newcommand{\as}{\mbox{$^{\prime\prime}$}}

\newcommand{\kms}{\mbox{km s$^{-1}$}}
\newcommand\cmc{\mbox{cm$^{-3}$}}
\newcommand\cmsq{\mbox{cm$^{-2}$}}
\newcommand{\um}{$\mu$m}
\newcommand{\cm}{cm$^{-1}$}

\def\lsim {$\rlap{\raise.4ex\hbox{$<$}}\lower.55ex\hbox{$\sim$}\,$}

\newcommand{\iso}{\mbox{\it ISO}}

\newcommand\ir{infrared}

\newcommand\mir{mid-infrared}

\newcommand{\hhcol}{\mbox{N$_{\rm H_2}$}}

\newcommand{\hh}{\mbox{{\rm H}$_2$}}

\newcommand{\hho}{H$_2$O}
\newcommand{\nhhh}{NH$_3$}
\newcommand{\oco}{CO$_2$}
\newcommand{\coo}{CO$_2$}

\newcommand{\hcch}{C$_2$H$_2$}

\newcommand{\ihcch}{$^{13}$C$^{12}$CH$_2$}
\newcommand{\chhh}{CH$_3$}
\newcommand{\chhhh}{CH$_4$}

\newcommand{\methanol}{CH$_3$OH}
\newcommand{\ocn}{OCN$^{-}$}
\newcommand{\ethane}{C$_2$H$_6$}
\newcommand{\chh}{CH$_2$}

\newcommand{\nufi}{$\nu_5$}
\newcommand{\nufo}{$\nu_4$}
\newcommand{\nuto}{$\nu_2$}
\newcommand{\nuff}{$\nu_4$~+~$\nu_5$}

\newcommand{\ngc}{NGC~7538}
\newcommand{\irs}{IRS~1}

\shorttitle{MIR Spectroscopy of \ngc\ IRS~1}
 \shortauthors{Knez et al.}

\begin{document}

\title{High Resolution Mid-Infrared Spectroscopy of NGC 7538 IRS 1:
Probing Chemistry in a Massive Young Stellar Object}

\author{Claudia Knez\altaffilmark{1,2,3}, John H. Lacy\altaffilmark{1,3},
Neal J. Evans, II\altaffilmark{1,3}, Ewine F. van
Dishoeck\altaffilmark{4}, Matthew J. Richter\altaffilmark{3,5}}

\altaffiltext{1}{Department of Astronomy, University of Texas at
Austin, Austin, Texas  78712 }

\altaffiltext{2}{Department of Astronomy, University of Maryland,
College Park, Maryland 20742; Email: claudia@astro.umd.edu}

\altaffiltext{3}{Visiting Astronomer at the Infrared Telescope
Facility, which is operated by the University of Hawai'i under
contract with the National Aeronautics and Space Administration.}

\altaffiltext{4}{Leiden Observatory, PO Box 9513, 2300 RA Leiden,
The Netherlands}

\altaffiltext{5}{Department of Physics, University of California,
Davis, One Shields Ave, Davis, California 95616}

\begin{abstract}

We present high resolution (R = 75,000--100,000) \mir\ spectra of the
high-mass embedded young star \irs\ in the \ngc\ star-forming
region. Absorption lines from many rotational states of \hcch,
\ihcch, \chhh, \chhhh, \nhhh, HCN, HNCO, and CS are seen. The gas
temperature, column density, covering factor, line width, and
Doppler shift for each molecule are derived.  All molecules were fit
with two velocity components between --54 and --63 \kms.  We find
high column densities ($\sim 10^{16}$ \cmsq) for all the observed
molecules compared to values previously reported and present new
results for \chhh\ and HNCO. Several physical and chemical models
are considered. The favored model involves a nearly edge-on disk
around a massive star. Radiation from dust in the inner disk passes
through the disk atmosphere, where large molecular column densities
can produce the observed absorption line spectrum.

\end{abstract}


\section{Introduction}
\label{sec:intro}

The \ngc\ (S158) molecular cloud, located in the Perseus arm at a
distance of 2.8 kpc \citep{ung00}, contains many protostellar
objects, making it a good candidate to study different stages of
star formation. Furthermore, there seems to be a progression in
evolution from northwest (the visible H~II region, \ngc, and IRS~2,
an only moderately obscured compact HII region) to southeast (IRS 9,
a deeply embedded core) \citep{elmegreen77,campthomp84}. The near-
and \mir\ sources IRS~1, 2, and 3 all lie near the northwest end of
the star-forming region. Of these, \irs\ is the brightest \mir\ and
radio continuum source \citep{martin73,wynnwilliams74,willner76}.
Its exciting star has a luminosity of $L > 8 \times 10^{4}~
L_{\odot}$, which corresponds to a ZAMS spectral type earlier than
O7.5, and an ionizing photon flux of $\phi > 10^{48} ~{\rm s}^{-1}$
\citep{werner79,lugo04}.

\irs\ is embedded in a dense molecular cloud. Millimeter continuum
emission is seen from the surrounding dust
\citep{scoville86outflows}.
\citet{pratap89} mapped the HCO$^+$ and HCN emission from within
$\sim$1\am\ of \irs\ with 3\as\ resolution. They found evidence for
a shell-like structure on scales of 10--20\as\ around \irs.
\citet{wilson83} observed \nhhh\ absorption toward \irs, and
\citet*{henkel84} mapped the absorption with the VLA.  The
absorption traces warm gas with a high column density of \nhhh\ and
is centered near $-60$~\kms.  The emission, on the other hand, is
centered near $-56.5$~\kms.  The systemic velocity for this source
is $\sim 57$~\kms\ \citep{vdt00}. More recently, \citet{zheng01}
mapped \nhhh\ emission from a more extended region in \ngc. This
emission probably probes the outer envelope as indicated by the
colder temperature and the smaller column density. Using both
single-dish and interferometer observations of various molecular
tracers, \citet{vdt00} studied the density and temperature structure
of both the cold outer envelope and the warm inner material
(240--72000 AU).  The inner warm region is characterized by
temperatures of a few hundred K.

Outflows are often present toward protostars, and \ngc\ \irs\ is no
exception. \citet{campbell84} observed \irs\ with the VLA at 5 and
15 GHz. She found a pair of very compact lobes of continuum
radiation, separated in declination by 0.2\as, with emission
extending out to $\pm$2\as. Her preferred model involves a bipolar
ionized outflow from a late O star, collimated by a core of dense
gas extending from $<$65~AU to $>$25,000~AU.  Further evidence of
outflows was found when \citet{gaume95} observed \irs\ with the VLA
in the H66$\alpha$ line and the 22 GHz continuum.  Their continuum
image is similar to that of Campbell, but shows additional
structure. They found broad recombination line emission
in the two lobes, with a
minimum line-to-continuum ratio, which they attribute to high
electron density, between the lobes. They propose a model involving
a high velocity stellar wind interacting with photo-evaporating
knots of neutral gas.

More recently, \citet{lugo04} have modeled the radio continuum
observations of \irs\ as due to a wind produced by photo-evaporation
of a circumstellar disk with a radius of 500 AU exposed to UV
radiation from the central O star.  Maser line emission is seen from
\hho\ \citep{kameya90}, OH \citep{dickel82}, H$_2$CO
\citep{hoffman03}, \methanol\ \citep{menten86}, and \nhhh\
\citep{madden86}. \citet*{minier98} used VLBI observations of the
methanol masers to infer the presence of a nearly edge-on rotating
disk traced by one of the maser clusters toward \irs. The central
velocity of this cluster, which is the brightest cluster near IRS 1 in both
\methanol\ and \nhhh\ masers, is --56.2~\kms. A second cluster of
masers, 0.25\as\ to the south, is at --61.0 \kms. The disk masers
lie along a line centered near the gap between the lobes of the
free-free continuum radiation, although they are aligned along a
position angle $\sim$30\degree\ from the symmetry plane.
\citet{pestalozzi04} made a detailed model of maser emission from a
disk around \irs, supporting the suggestion of \citeauthor{minier98}
They concluded that the maser emission comes from disk radii of
$\sim 290 - 750$~AU.  However, more recently, \citet{debuizer05}
show that the \methanol\ masers may trace knots in the outflow
instead.  Also, \citeauthor{debuizer05} identify a disk-like
structure perpendicular to the outflow direction in mid-infrared
continuum.  The extent of the infrared emission is $\sim$450 AU.

Infrared absorption by dust grains and their icy mantles has also
been observed toward \irs. \citet{willner76} and \citet{willner82}
observed 9.7~\um\ silicate absorption and absorption by various
ices, including \hho.  The ice bands toward \irs\ are relatively
weak, especially in comparison with \ngc\ IRS~9, which must lie in a
colder region of the \ngc\ cloud. \citet{gibb04} observed the
complete 2.5-20~\um\ region with the Short-Wavelength Spectrometer
(SWS) on the {\it Infrared Space Observatory}, \iso. \hho, CO, and
\oco\ ices are clearly present, as well as other less securely
identified features. Toward \ngc\ IRS~9, $\sim$1\arcmin\ southeast
of \irs, a much richer ice spectrum is seen, with CH$_3$OH, XCN
(probably OCN$^-$), and \chhhh\ also confidently identified, and
OCS, H$_2$CO, HCOOH, and \nhhh\ likely present.

Gas-phase absorption in the infrared has been observed toward \irs\
by several groups. \citet{mitchell90} observed absorption by
$^{12}$CO and $^{13}$CO in their 5~\um\ $v=1-0$ bands. They did not
spectrally resolve the lines, and they used a curve of growth
analysis assuming pure absorption by Gaussian lines with FWHM of
8~\kms\ to derive the gas temperature and column density from the
less-saturated $^{13}$CO lines. They found two absorbing components:
cold gas at 25~K, and warm gas at 176~K. A gas density of $n_H >
10^{6}$~\cmc\ was required to maintain the population of the
high-$J$ levels in the warm gas. Using the \iso-SWS,
\citet{lahuis00} detected warmer gas than the CO observed by
\citeauthor{mitchell90}. They observed the $\nu_5$ band of \hcch\ at
13.7~\um\ and the $\nu_2$ band of HCN at 14.0~\um. With the SWS
resolution of $\sim$1800, they were not able to resolve individual
line shapes, and were most sensitive to the blended lines of the Q
branches, but they were able to derive temperatures and column
densities from the shapes of the Q branches and the depths of the
$v=1-0$ fundamental and $v=2-1$ hot band Q branches. They derived
$T$ = 800~K and $N = 0.8 \times 10^{16}$ \cmsq\ for \hcch, and
$T$~=~600~K and $N = 1.0 \times 10^{16}$ \cmsq\ for HCN.
\citet{boonman03} derived $T$ = 500~K for \hcch\ from an updated
reduction of the spectra. \citet{boonman03} and \citet{boonman03h2o}
studied gas absorption from \coo\ and \hho\ using \iso\ data.  They
found enhanced abundances toward the inner warm material compared to
the cold envelope indicating that grain mantles are sublimating and 
enriching the gas-phase chemistry close to the
protostar.  Gas-phase \chhhh\ was also detected toward IRS~9
\citep{lacy91, boo04meth} but not \irs.

The wealth of observational data available suggest the following
physical scenario for \ngc\ \irs.  The massive young star is
surrounded by a massive cold envelope on scales $\sim$ 72000 AU
\citep{vdt00}.  At scales below 1000 AU, the temperature increases
to $\geq$ 100 K. Inside this radius, there is evidence for a disk
\citep{debuizer05} and small scale knotty outflows \citep{gaume95},
both of which can be affected by the UV and X-ray radiation from the
protostar  \citep[e.g.,][]{stauber04, stauber05}.

In this paper, we present high resolution \mir\ spectroscopy of
\ngc\ \irs\ showing a rich absorption spectrum containing lines from
{\it seven} molecules: \hcch, HCN, \chhhh, \chhh, \nhhh, HNCO, and
CS. Mid-infrared absorption is much more sensitive to the inner warm
gas than radio observations.  Previous infrared studies suffered
from limited spectral resolution. Section \ref{sec:obs} describes
the observations.  We then present the model used to derive the
column densities and temperatures for the various species in section
\ref{sec:fit} and describe the results for each of the molecules in section
\ref{sec:results}.
Subsequently, in section \ref{sec:mod}, we discuss the
possible scenario in which we are probing chemistry in a
circumstellar disk. In section \ref{sec:conc}, we provide some
concluding remarks.

\section{Observations}
\label{sec:obs}

\subsection{Observations and Data Reduction}

The observations were made with TEXES, the Texas Echelon Cross
Echelle Spectrograph \citep{lacy02}, on the NASA Infrared Telescope Facility (IRTF)
in 2001 June and November, 2002 September and December, and 2005
December.  TEXES is a high-resolution cross-dispersed spectrograph operating at
\mir\ wavelengths between 5 and 25~\um. It achieves
a spectral resolution of $R$ = 75,000-100,000 ($\Delta v =
3-4$~\kms) shortward of 14~\um\ with a slit width of $\sim$1.5\as\ .
(Of the observations presented here, only the spectra near 11~\um ,
which used a 1\as\ slit, achieved the higher resolution.) 
The spatial resolution along the North-South oriented slit is $\sim$1\as , slightly
larger than the diffraction limit of the IRTF. The spectral and
spatial sampling by the $256 \times 256$ pixel Si:As detector array
are 1.0~\kms\ and 0.35\as , respectively.  At each
spectral setting, 5-10 orders of the high resolution echelon grating
are recorded, giving a spectral coverage of $\Delta\lambda_{tot}/\lambda~\sim~$0.5\%. At
wavelengths shortward of 11~\um\ full order widths are observed, giving
continuous spectral coverage, but longward of 11~\um\ the order
width exceeds the array width, leaving gaps in the observed spectra.  The location of the 
gaps can be shifted in order to optimize line coverage.  However, since the gaps are small
and, at 13 \micron, there are many telluric lines, 
it would be impractical to repeat observations with small shifts in order to get complete 
spectral coverage for each setting.

Much of the data were taken in high water vapor conditions,
but the observing conditions did not seriously affect the spectra
except for increased noise, especially on atmospheric lines.
Observations of \ngc\ \irs\ were interspersed with observations of a
bright asteroid (usually Ceres ($> 500$ Jy) , but also Hygeia ($\sim 150$ Jy), which served as a
comparison source for removal of telluric absorption features. The
telescope was nodded by 3--5\as, moving the source along the
spectrograph entrance slit to allow subtraction of background
emission. At the beginning of each set of nodded observations an
ambient temperature blackbody was observed for flat-fielding.  The first 
nods were used to peak up on the source by maximizing the 
throughput of the slit.  The nod pairs taken during peak up were not included
in the final sum.

Data reduction followed the standard TEXES procedure \citep{lacy02}
of first subtracting the two nod position spectra, flat-fielding
with the blackbody-sky difference echellogram  \citep{lacy89irsh}, and interpolating over
spikes and bad pixels. Optimally-weighted point-source spectra were
then extracted from the echellograms. The spectra were linearized in
ln$(\nu)$ based on the known optical distortions, and absolute
wavenumber calibration was obtained from telluric emission features
and corrected for the Earth's motion relative to LSR. The resulting
wavenumber scale is correct to within 1~\kms . The same procedure
was used for the asteroid spectra, and then the \irs\ spectra were
divided by the asteroid spectra with correction for differences in
airmass. With this procedure, telluric absorption lines as deep as
90\% can be removed, although of course the noise increases on
lines. However, broad response variations often remained, which were
removed by fitting the continuum in each echelon order to a quartic
polynomial and dividing. This procedure also sets the continuum in
each order to one.

\subsection{Line Identification and New Infrared Detections}

\begin{figure*}
 \begin{center}
 \epsfig{file=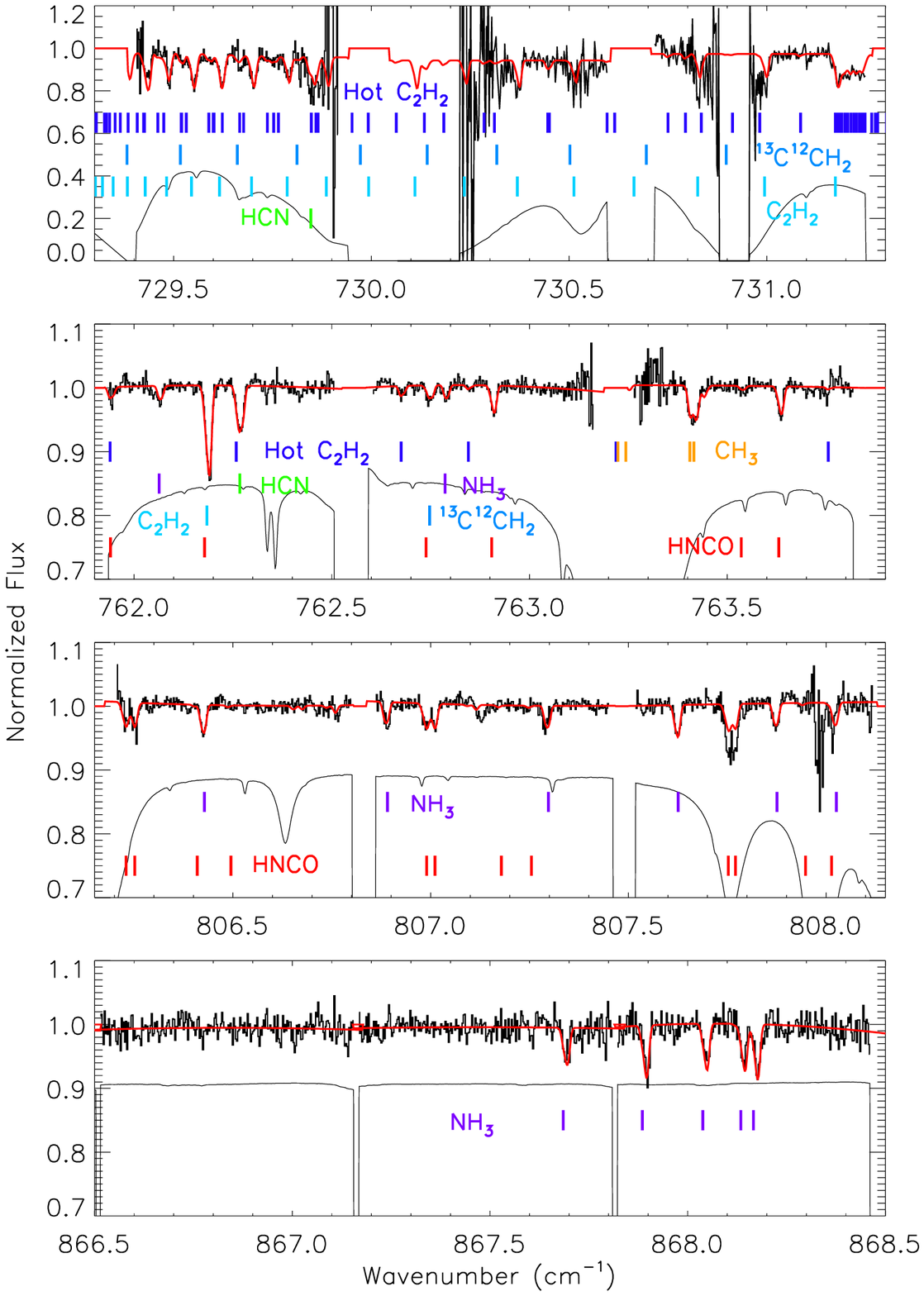,height=9.0in}
 \caption[Selected orders between 729 and 870 \cm]{Selected orders
(thick black) between 729 and 870 \cm\ are shown with the best fit
(red). Positions of lines considered in the fit are given: \hcch\
(light blue), \ihcch\ (blue), hot bands of \hcch\ (dark blue), HCN
(green), \chhh\ (orange), HNCO (red), and \nhhh\ (purple). The thin
black line indicates the atmospheric transmission.  The top panel
has a larger scale for the normalized flux (y-axis) to show the
lower atmospheric transmission.  Note that there are gaps between
orders.
 \label{fig:twelveum} }
 \end{center}
\end{figure*}

\begin{figure*}
\begin{center}
\epsfig{file=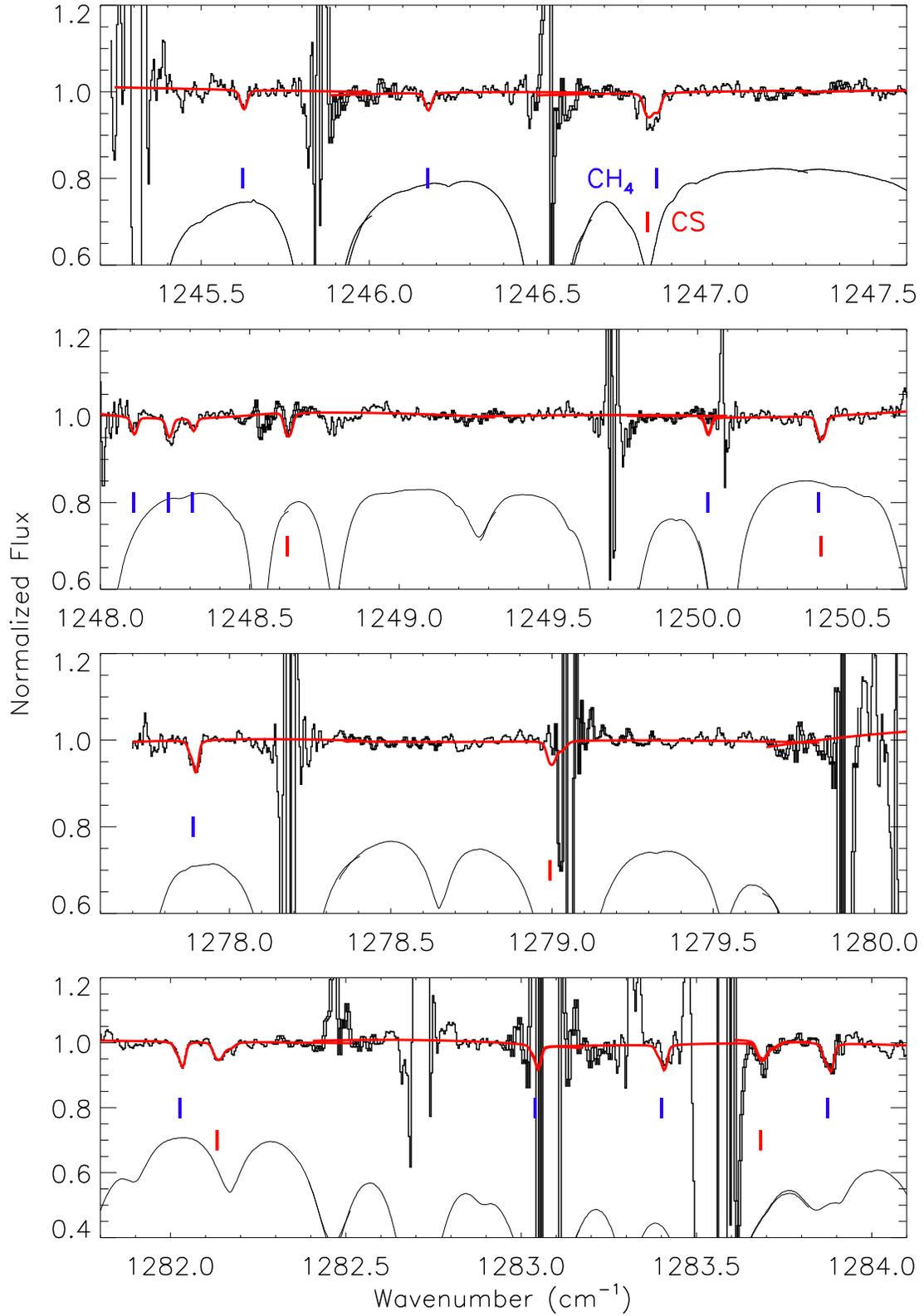, height=9.0in} \caption[Selected
orders between 1245 and 1285 \cm]{Selected orders (thick black)
between 1245 and 1285 \cm\ are shown with the best fit (red).
Position of lines considered in the fit are given: \chhhh\ (blue)
and CS (red). The thin black line indicates the atmospheric
transmission.  In this region there are no gaps between orders.
Instead the orders overlap where the black line is darker.
\label{fig:eightum} }
\end{center}
\end{figure*}

\begin{figure}
 \begin{center}
 \epsfig{file=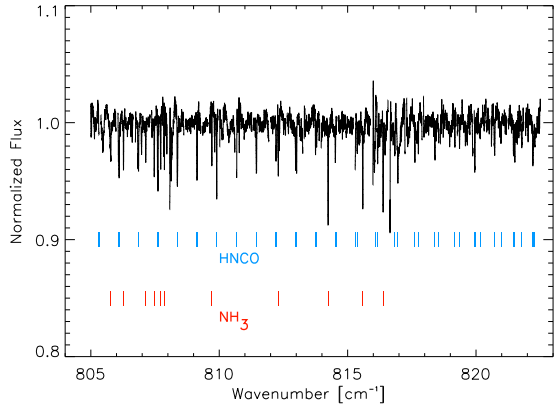, width=3.2in}
 \caption[Forrest of HNCO and \nhhh\ lines toward \ngc\ \irs]
{This figures shows a portion of spectrum containing some \nhhh\
lines as well as wealth of HNCO lines. These spectral settings led
to the identification of HNCO.  Figure \ref{fig:hncoblowup} shows a
small portion of this spectrum.
 \label{fig:hncoall}}
 \end{center}
\end{figure}

\begin{figure}
 \begin{center}
\epsfig{file=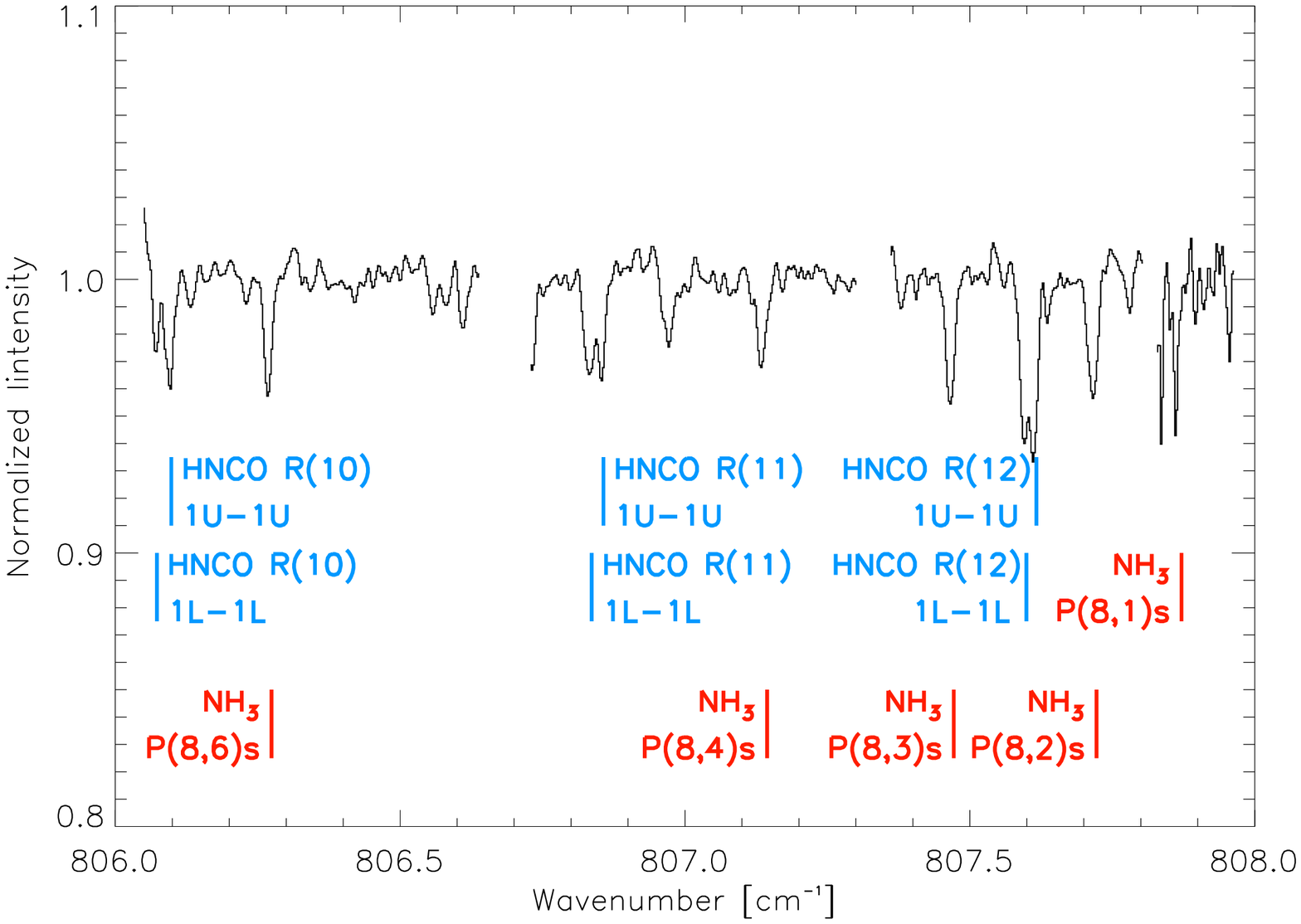, width=3.2in}
 \caption[Blow-up of HNCO and \nhhh\ lines]{This figure shows the profiles of several HNCO and \nhhh\
lines in the spectral region shown in Figure \ref{fig:hncoall}. The
two HNCO sub-bands shown are R-branch transitions where the K = 1
state is split into upper (U) and lower (L) levels.  Because of the
close separation the sub-bands, the lines can overlap for a range of
J.  See Appendix A for more details on transition rules for HNCO.
 \label{fig:hncoblowup}}
 \end{center}
\end{figure}

Ten spectral settings were observed in the 728--820~\cm\
(13.7--12.2~\um  ) region, two in the 860--930 \cm\ (11.5--10.7~\um
) region (see Figure \ref{fig:twelveum}), and three in the
1240--1312~\cm\ (8.0--7.6~\um ) region (see Figure
\ref{fig:eightum}). A total of 110 echelon orders, or $\sim$70 \cm ,
were observed. The original observations were meant to study the
line profiles of \hcch\ and HCN, which had previously been detected.
In addition to the prominent fundamental band lines of \hcch\ and
HCN, many \hcch\ hot band lines were detected in our spectra
including lines from the \nuff~--~\nufo\ and the 2\nufi~--~\nufi\
bands. After the hot bands were identified, unidentified lines still
remained, some of which were double dipped. The double lines turned
out to be \chhh\ lines which are split due to spin-rotation
interactions (see Appendix). This is the first detection of \chhh\
toward dense gas and the first ground-based detection of \chhh.  Previously, 
\chhh\ had only been observed with \iso\ toward the Galactic center \citep{feucht00}. In
efforts to detect ethane, \ethane, at 810-820 \cm, many lines were
detected, though none corresponded to \ethane. Some were due to
\nhhh, but many lines remained unidentified (see Figures
\ref{fig:hncoall} and \ref{fig:hncoblowup}). Based on their
separation we were able to identify the lines as due to HNCO.
Interstellar HNCO is a well known hot core molecule at radio
wavelengths \citep{zinchenko00}, but it had never been seen at
infrared wavelengths. In total, five molecules were observed in the
728--820~\cm\ region: \hcch\ (including \ihcch), HCN, HNCO, \chhh ,
and \nhhh. At higher frequencies, \chhhh, \hcch\ and CS were
observed in the 1240--1320 \cm\ region. \chhhh\ observations from
the ground are possible for this source because of the large Doppler
shift with respect to the telluric lines.

Table \ref{sumlines} gives a summary of the detected lines. At least
5 lines were observed for each molecule, with some molecules like
\hcch\ (including isotopic lines and hot bands), and \nhhh\ having
more than 40 lines detected. However, the number of HNCO lines
dominates over lines from the other molecules, with over 100 lines
detected. Table \ref{linelist} (full version is available online
only) lists all the individual lines detected in our spectra.  A
selection of spectra including all observed molecules are shown in
Figures \ref{fig:twelveum} and \ref{fig:eightum}, along with fits
described in \S \ref{sec:fit}. For most molecules the line strengths
were taken from the GEISA03
database\footnote{http://ara.lmd.polytechnique.fr}. The \chhh\ band
strength was obtained from \citet{wor89} and the HNCO oscillator strength
was obtained from \citet{low01}. The CS oscillator strength was taken from
\citet{bot85}.

\begin{deluxetable}{lcc}
 \tablecaption{{Summary of Observed Lines}}
 \tablewidth{0pt}
\tablehead{ \colhead{Molecule}      &
    \colhead{Total Number of}    &
    \colhead{Band}           \\ 
    \colhead{}          &
    \colhead{Detected Lines}          &
    \colhead{}
} \startdata

\hcch       &   25  &   \nufi       \\ 
\hcch$^a$   &   2   &   \nuff       \\ 
\hcch$^b$   &   20  &   \nuff--\nufo \\ 
\ihcch$^c$  &   7   &   \nufi       \\ 
HCN         &   10  &   \nuto       \\ 
\nhhh       &   64  &   \nuto       \\ 
HNCO        &  125  &   \nufo       \\ 

\chhh       &   11  &   \nuto       \\ 
\chhhh      &   12  &   \nuto/\nufo\ dyad       \\ 
CS          &   6   &   \nuto               

\label{sumlines}
\enddata
\tablecomments{$a$.  These lines were observed at 8 \micron. \\
$b$. Lines in Q-branch at 731 \cm.  See top panel of Figure
\ref{fig:twelveum}. \\
$c$. Only unblended lines are reported here.}
\end{deluxetable}

\begin{deluxetable}{lcccc}
\tablecolumns{5} 
\tablecaption{{Observed Lines}}
 \tablewidth{0pt} \tablehead{
\colhead{Molecule}      &
    \colhead{Line}    &
    \colhead{Wavenumber}        &
    \colhead{$E_J$}          &
    \colhead{Band}     \\
    \colhead{}          &
    \colhead{}          &
    \colhead{\cm}   &
    \colhead{(K)}          &
    \colhead{}
} \startdata

\hcch\      & Q(5)  &  729.289 &    35.30      &  \nufi   \\
\hcch\      & Q(6)  &  729.343 &    49.42      &  \nufi   \\
\hcch\      & Q(7)  &  729.406 &    65.89      &  \nufi   \\
\hcch\      & Q(8)  &  729.477 &    84.71      &  \nufi   \\
\hcch\      & Q(9)  &  729.558 &   105.89      &  \nufi   \\
\hcch\      & Q(10)  &  729.648 &   129.41      &  \nufi   \\
\hcch\      & Q(11)  &  729.747 &   155.29      &  \nufi   \\
\hcch\      & Q(14)  &  730.096 &   247.02      &  \nufi   \\
\hcch\      & Q(15)  &  730.230 &   282.30      &  \nufi   \\
\hcch\      & Q(16)  &  730.373 &   319.93      &  \nufi   \\
\hcch\      & Q(18)  &  730.686 &   402.22      &  \nufi   \\
\hcch\      & Q(19)  &  730.855 &   446.89      &  \nufi   \\
\hcch\      & Q(20)  &  731.034 &   493.90      &  \nufi   \\
\hcch\      & Q(26)  &  732.287 &   825.20      &  \nufi   \\
\hcch\      & Q(27)  &  732.526 &   888.62      &  \nufi   \\
\enddata
\label{linelist}

\tablecomments{Only lines that were individually detected at the
3-$\sigma$ level are included in this list. Full line list is
available online only.}

\end{deluxetable}

\section{Spectrum Fitting}
\label{sec:fit}


To derive physical parameters of the absorbing gas, model spectra
were fitted to the observations. For this purpose, the data were
divided into two groups. All data at wavenumbers between 700 and
1000 \cm\ (10.7--13.7 \micron) were fitted simultaneously, and
data at 1240--1280 \cm\ (near 8 \micron) were fitted separately. The
Marquardt fitting procedure \citep{bev03} was used, which minimizes
$\chi2$, the summed squared deviation normalized by the squared
noise, between the data and the model.

\begin{figure}
 \begin{center}
 \epsfig{file=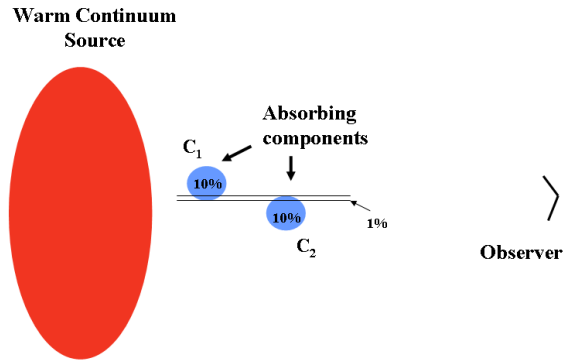,angle=270, width=3.2in}
 \caption[Illustration of two component model]
{This figure shows two absorbing components between the continuum
source and the observer.  Each component covers 10$\%$ of the
continuum source.  The overlap of the two absorbing regions is given
by the product of their covering factor.  In this case, only 1$\%$
of the continuum source is covered by both components.
 \label{fig:comp}}
 \end{center}
\end{figure}

Fitting was first attempted with a single component for each
molecule, but the residuals suggested that a better fit could be
obtained with multiple velocity components.  The model that we used
assumed that each molecule is found in one or two absorbing
components. Each component is described by its Doppler shift,
$V_{\rm LSR}$, its Gaussian line width ($1/e$ half width = (2kT/m)$^{1/2}$), $b$, the
molecule's column density, $N$, the rotational temperature of the
gas, $T$, and the covering factor, $C$, (the fraction of the
background continuum source covered by the component). The fitted
column density is the average over the partially covered source, so
the column density in the covered portion would be $N/C$. A covering
factor of, e.g., 10\% means that lines saturate at 90\% of the
continuum flux and the column density in the absorbing component
is ten times the average over the continuum source.
The effect modeled in this way could also result from
veiling (continuum emission from foreground or surrounding
material) or from re-emission in the lines by the absorbing
molecules.  The components were assumed to overlap by the
products of their covering factors (see Fig. \ref{fig:comp}). The
following equation was used to calculate the observed transmission:

\begin{equation}
I_{obs} = I_0 (1 - C_1 (1 -  e^{-\tau_1({\nu})})) (1 - C_2 (1 -
e^{-\tau_2({\nu})})),
\end{equation}

\noindent where $C_1$ and $C_2$ are the covering factors for the two
components.  This results in a greater absorption in saturated lines
than would result if the optical depths were added first and then
the transmission spectrum was calculated from the optical depth
spectrum. This approach was chosen because it gave a better fit to
the data than adding the optical depths first and assuming the same
covering factor, but it assumes that different components absorb
along different lines of sight to a partially covered continuum
source, which may not be the case.

A frequency correction was allowed for each spectral setting
observed, to correct for errors in wavelength calibration. With only
one exception, the correction was less than a 1~\kms\ Doppler shift.
In addition, a broadening of the instrumental resolution was
permitted in the fitting to allow for imperfect internal instrument
focusing, or possibly a `macroturbulent' broadening in the absorbing
gas. The fitting program chose a resolution a factor of 1.2 greater
than that derived from gas cell data. A constant frequency
resolution (as opposed to a constant Doppler resolution) was used in
each of the two fitted regions. This represents well the improving
resolving power toward shorter wavelengths in the 10.7-13.7~\um\
region. In addition to these parameters, the continuum, slope, and
curvature of each echelon order were varied, to allow correction of
the baseline fitting done during data reduction. The continuum
fitting and frequency correction required about 300 free, but rather
easily determined, parameters. Fewer parameters are needed to
determine the physical conditions such as temperature, column
density, line width and covering factor:  70 parameters for the
10.7-13.7~\um\ spectra and 21 for the 8~\um\ region.  For the
10-13~\micron\ region, we have over 3000 points to constrain the 70
parameters of interest if we characterize the constraining points by
the number of lines times the number of pixels for each line.

For \hcch\ and HNCO, lines of several bands were observed. For
\hcch, in addition to lines of the Q and R branches of the $\nu_5$
fundamental, lines of several \nufo~+~\nufi~--~\nufi\ and
2\nufi~--~\nufi\ bands (involving absorption from the excited
$\nu_4$ and $\nu_5$ vibration states) were observed. The vibrational
temperatures describing the populations of the 
$\nu_4$ and $\nu_5$ states were included as free
parameters in the fit.  R-branch lines of the $\nu_5$ band of
\ihcch\ were also observed, and the \ihcch/\hcch\ abundance ratio
was allowed to vary, although it was kept the same in all the \hcch\
components.  Two \hcch\ lines of the \nuff\ band at 7.6~\micron\
were observed as well. For HNCO, lines from the \nufo\ band were
observed. HNCO is an only slightly non-linear molecule, and its
spectrum consists of a series of sub-bands resembling those of a
linear molecule, like \hcch. Its dipole moment oscillates
diagonally, giving it a-type sub-bands, with no change in the
angular momentum about its long axis and only P and R branches, and
b-type sub-bands, in which the angular momentum about its long axis
changes, and P, Q, and R branches are seen.  See Appendix for
further discussion of the HNCO spectrum. Since we did not have
laboratory data regarding the relative strengths of the different
sub-bands, we considered them to be caused by different species and
summed the abundances to obtain the HNCO abundance. The HCN lines
observed were from the $\nu_2$ bending mode. We included HCN
2\nuto\ -- \nuto\ lines in the fit.  Although no hot HCN lines were obvious
in the spectra, their inclusion lowered $\chi^{2}$ significantly
($\Delta \chi^{2} = 13$) and changed the best-fitting $N_{\rm HCN}$
noticeably.

Although it might not be apparent from the spectra, given the
shallowness of the lines, many of the detected lines are at least
moderately saturated, and some are very saturated. The lines only
appear weak because of the small covering factors (or emission
filling in the absorption lines) and to a lesser extent the small
line widths. This conclusion is very robust; the relative depths of
lines of different opacity requires substantial saturation.  This is
most apparent in the near equality of the depths of the ortho and
para lines in the \hcch\ Q branch, near 730 \cm\ (see top panel of
Figure \ref{fig:twelveum}). HCN in the --60~\kms\ component and
NH$_3$ in the --56.5~\kms\ component are so saturated that only
lower limits could be placed on their abundances until their
isotopomers were included in the fit to provide upper limit
constraints. By including isotopomers and intrinsically weak lines,
a wide enough range of line optical depths was observed to allow
meaningful constraints to be placed on the molecular abundances in
spite of the saturation.

The observed lines of CH$_3$ were in the R branch of the $\nu_2$
out-of-plane bending mode of this planar molecule. Each of these
lines is doubled by the interaction between the electron spin and the
molecular rotation (see Appendix). The CH$_3$ $\nu_2$ band center is
at 16.5~\um, whereas all observed lines were shortward of 13.5~\um,
with $J \geq$ 7, making us rather insensitive to cold CH$_3$. Lines
of the P and Q branches of the NH$_3$ $\nu_2$ band were observed.
There was no evidence of $^{15}$\nhhh\ lines, but they were included
in the fit to constrain the \nhhh\ abundance, as was done with
H$^{13}$CN. In the 8~\um\ region, which was fitted separately, lines
of the $\nu_2 / \nu_4$ dyad of CH$_4$ were observed. Because of the
strong absorption by low-$J$ lines of CH$_4$ in the Earth's
atmosphere, most of the observed lines originate at $J > 4$.  We
were able to observe the R(0) line due to the favorable Doppler
shift of the source ($\sim -57$~\kms).  The R(0) line helps
constrain the temperature derived from the high $J$ lines. We also
detected and fitted absorption by CS.

The derived component parameters are given in Table
{\ref{tabparms}}. Uncertainties for the parameters, given in
parentheses, are three times the square roots of the diagonal
elements of the error matrix, which are uncertainties allowing all
other parameters to vary. The noise in the fitted spectra is assumed
to be purely photon statistical noise. With this noise estimate, the
reduced $\chi^{2} \approx 1.6$, indicating that non-statistical
noise sources contribute moderately, or perhaps that a different
model for the line profiles is justified. Including non-statistical
noise, the uncertainties given might reasonably be taken to be 95\%
confidence intervals.  The larger fractional errors are asymmetric, and 
would be more symmetric in the log of the parameters.  For column densities, 
this is a result of the non-linearity of the curve of growth for moderately saturated lines.
Column densities with errors greater than their values are not consistent
with zero.  For unsaturated lines, the derived covering factors are very
uncertain because equivalent widths depend only on the product of the
covering factor and the column density.  For example, 
the column densities of HNCO and \chhh\ have smaller fractional
uncertainties than those of \hcch\ and HCN, whereas the covering
factors have smaller uncertainties for \hcch\ and HCN. 
All of the fitted molecules are detected with very high
confidence.

\begin{deluxetable*}{lccccccccccc}
\tablecolumns{12}
 \tablewidth{0pt}
\tablecaption{Results from the model fit to the spectra}
\tablehead{
  \colhead{} & \multicolumn{5}{c}{Component 1}    & \colhead{} &
  \multicolumn{5}{c}{Component 2} \\
  \cline{2-6} \cline{8-12} \\
  \colhead{Molecule}            &
  \colhead{$V_{\rm LSR}$}      &
  \colhead{$b$}         &
  \colhead{$T_{ex}$}            &
  \colhead{$N(X)$}                &
  \colhead{C}                   &
  \colhead{}	&
  \colhead{$V_{\rm LSR}$}      &
  \colhead{$b$}         &
  \colhead{$T_{ex}$}            &
  \colhead{$N(X)$}                &
  \colhead{C}                   \\
  \colhead{}                    &
  \colhead{(\kms)}                &
  \colhead{(\kms)}              &
  \colhead{(K)}                 &
  \colhead{($10^{16}$ \cmsq)}   &
  \colhead{($\times 100$)}			&
  \colhead{}                         &
  \colhead{(\kms)}                &
  \colhead{(\kms)}              &
  \colhead{(K)}                 &
  \colhead{($10^{16}$ \cmsq)}   &
  \colhead{($\times 100$)}
} \startdata
\hcch  & --55.7 (0.3)   & 0.6 (0.1)  &  225 (20)  & 3.0 (0.6)  & 5.9 (0.6)  & &  --59.4 (0.3)   & 0.6 (0.1)  &  191 
(10)  & 2.8 (0.2)  & 23.7 (1.0) \\
HCN    & --56.3 (0.3)   & 0.8 (0.1)  &  256 (30)  & 5.6 (2.8)  & 8.4 (0.9)   & & --60.0 (0.3)   & 1.0 (0.4)  &  456 
(127) & 1.3 (1.2)  & 68.4 (49.7)  \\
HNCO   & --57.2 (0.3)   & 1.8 (0.2)  &  319 (27)  & 0.4 (0.3)  & 3.8 (0.3)   & & --60.2 (0.2)   & 0.7 (0.4)  &  
171 (79)  & 0.1 (0.3)  & 97.0 (11)  \\
\chhh  & --54.2 (0.4)   & 1.8 (0.6)  &  258 (17)  & 1.7 (0.2)  & 97.0 (15) & & --62.8 (0.5)   & 1.5 (0.9)  &  927 
(161) & 0.6 (0.1)  & 97.3 (18)   \\
\nhhh  & --57.3 (0.3)   & 0.9 (1.0)  &  278 (31)  & 5.2 (1.5)  & 8.9 (1.2)  & & --60.1 (0.2)   & 0.4 (0.1)  &  248 
(32)  & 2.8 (0.8)  & 15.8 (3.6)   \\
\chhhh & --56.3 (0.3)   & 0.2 (0.04) &  674 (360) & 20 (8)     & 5.8 (1.9)  &  & --60.1 (0.3)   & 0.6 (0.3)  &  
668 (250) & 16 (13)    & 99 (13)   \\
CS     & --55.2 (0.4)   & 1.0 (0.3)  &  224 (173) & 0.02 (0.03)& 100 (8)      &  & --59.4 (0.5)   & 0.7 (0.3)  &  
249 (100) & 7.0 (3.0)  &  5.5 (0.3)   \\
 \label{tabparms}
\enddata
\tablecomments{The column densities listed are $N_{tot}\times C$. }
\end{deluxetable*}

\section{Results}
\label{sec:results}

For most molecules, we found two components with Doppler shifts of
${V_{\rm LSR} = -56.4 \pm 0.3}$~\kms\ and $-59.7 \pm 0.3$~\kms, with
widths of $b~=~1-2$~\kms\ (see Table \ref{tabparms}). Most of the
component temperatures were in the range 200-400~K and covering
factors were typically 0.04-0.08, with \hcch\ requiring somewhat
larger values of 0.06 and 0.18.  HNCO has somewhat broader component
line widths but other parameters are similar. \chhh\ has the
broadest component line widths, with a greater velocity spacing
between the two components and the hottest component.  Both CS and
\chhh\ are fitted with much larger covering factors (consistent with
100$\%$) than were the other molecules.  However, these covering
factors are quite uncertain due to the weakness of the observed lines,
and the large values may not be significant.
The model spectra resulting
from the best fit parameters can be seen in Figures
\ref{fig:twelveum} and \ref{fig:eightum}.

In addition to agreeing among the different molecules (except
\chhh), the Doppler shifts agree with the two most commonly observed
values for molecules seen at radio wavelengths. This gives us some
confidence that the two velocity components, at --56.5 and
--60~\kms, are real, although whether they are two separate
components, or just a way of describing a non-Gaussian velocity
distribution is difficult to determine from the data. Generally, the
temperatures of the two components for a given molecule are not very
different, although a fit constraining them to be equal was
significantly worse than the fit allowing all temperatures to vary.
Line widths, column densities, and covering factors vary more. The
fit was very poor when they were constrained to be equal for all
molecules.

Note that our quoted column densities are averages over the lines of
sight to the continuum emitting material, or column densities in the
absorbing gas multiplied by the covering factors.
There are several possible interpretations for the covering factor parameters:
the continuum source could in fact be partially covered, foreground
or surrounding emission could veil the spectrum, or re-emission in the
lines could fill in the absorption.
In all of these cases, the column densities through the absorbing material
are given by N/C, rather than N, the average column density over the line of
sight in the partial covering interpretation.
Consequently, if we could determine the \hh\ column density along the
lines of sight on which we see absorption we should compare that to our
N/C values.

\subsection{\hh\ and CO From Previous Measurements}

Before discussing the results of our measurements of individual molecules,
we attempt to derive \hh\ and CO column densities from previous
measurements.  
The \hh\ column toward \ngc\ \irs\ has not been directly measured,
but it may be estimated from the extinction measurements of
\citet{willner76} with the assumption of a normal interstellar
gas-to-dust ratio. This method gives \hhcol\ = $6-9 \times 10^{22}$
\cmsq. If the continuum source is non-uniformly covered, this column
density represents an average over the source. Hence, it should
probably be compared to the values we derive for average column
densities, N, of the molecules we observe. However, the silicate
optical depth includes extinction from all material along the line
of sight to \ngc, most of which is unlikely to contain our
molecules.  This suggests that $6-9 \times 10^{22}$ \cmsq\ is an
overestimate of \hhcol\ in the relevant gas. On the other hand, we
suggest below that we may be observing absorption by gas in the
atmosphere of a protoplanetary disk, and dust in such a region could
be depleted by settling toward the disk midplane. This effect would
result in an underestimate of \hhcol\ from the extinction. We give
the ratios of our derived column densities to an \hh\ column density
of 7.5~$\times~10^{22}$ \cmsq\ in Table \ref{tab:abun}.

\begin{deluxetable}{lcc}
\tablecolumns{3}
 \tablecaption{Abundances with respect to CO and \hh}
 \tablewidth{0pt}
  \tablehead{
    \colhead{Molecule} &
    \colhead{N(X)/N(CO)*} &
    \colhead{N(X)/N(\hh)*} \\
    \colhead{}  &
    \colhead{10$^{-3}$} &
    \colhead{10$^{-7}$}
} \startdata
\hcch  & 5.7 &  7.7 \\
HCN    & 6.9 &  9.2 \\
HNCO   & 0.5 &  0.6 \\
\chhh  & 2.3 &  3.1 \\
\nhhh  & 8.1 & 10.8 \\
\chhhh & 36  &  48  \\
CS     & 7.0 &  9.3 \\
\label{tab:abun}
\enddata
\tablecomments{*N(CO) = 1.0$\times$10$^{19}$ \cmsq\ and N(\hh) =
7.5$\times$10$^{22}$ \cmsq.  See text for discussion of whether these
molecules are measured on the same lines of sight as the ones we observe.
We use $N$ listed in Table \ref{tabparms}. }

\end{deluxetable}

It may be preferable to compare our column densities to the column density
of CO, which has been observed in absorption toward \ngc\ \irs\ by
\citet{mitchell90}.  They obtain
a column density of $1.5 \pm 0.3 \times 10^{17}$~cm$^{-2}$ for
$^{13}$CO in cold gas at 25~K and $1.4 \pm 0.1 \times
10^{17}$~cm$^{-2}$ in warm gas at 176~K. We would not be sensitive
to their cold gas. Presumably their warm gas includes both of our
components, although our temperatures are generally higher than
theirs. If we assume $^{12}$CO/$^{13}$CO = 72,
using the $^{12}$C/$^{13}$C ratio we derived from our \hcch\ observations,
we derive the abundances of our molecules relative to CO listed in
Table \ref{tab:abun}.  (Note that \citeauthor{mitchell90} use
$^{12}$C/$^{13}$C = 89.) However, we should note that we have
measured one spectral setting including CO lines, while searching
for OCS absorption near 5~\um. The noise is rather high, but it is
apparent that the $^{13}$CO lines are not well described by
moderately saturated Gaussians, as assumed by
\citeauthor{mitchell90}. The $^{13}$CO lines are deeper and broader
than the lines of any of our other molecules, and have prominent
blue-shifted shoulders probably tracing an outflow. We suspect that
the lines are more saturated, and the CO column density larger than
\citeauthor{mitchell90} concluded. The different shapes and greater
depths of the CO lines than our \hcch\ lines also indicates that CO
probes different gas. Likely, \hcch\ and the other molecules we
observe in the \mir\ are found only in unusual regions, whereas CO
is distributed along the entire line of sight.

In general, we conclude that the ratios of column densities to those of
either \hh\ or CO given in Table \ref{tab:abun} are quite uncertain
because the different column density tracers may be sensitive to different
components along the line of sight.
The column densities of the different molecules we observe should be
much more comparable.

\subsection{\hcch\ and HCN}
\label{sec:c2h2}

\begin{figure}
 \begin{center}
 \epsfig{file=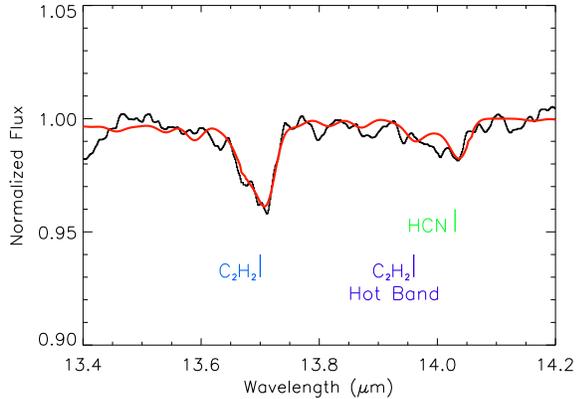, width=3.2in}
 \caption[Our fit compared to \iso\ spectrum]{The spectrum of NGC~7538
IRS~1 as seen by \iso\ is shown in black. A fit to the \hcch\ and
HCN fundamental Q-branches as well as a hot band Q-branch of \hcch\
is shown in grey.  The parameters are set from the TEXES
high-resolution observations.  The models gives a good fit to the
\iso\ spectrum. Color version is available online.
 \label{fig:isofit}}
 \end{center}
\end{figure}

Since our initial search for molecules began with \hcch\ and HCN,
the first models only included those two molecules.  In these
models, we derived similar temperatures and column densities as
derived from \iso\ observations \citep{lahuis00, boonman03}.  Once
we include absorption from other molecules, HNCO in particular, the
temperatures for HCN and especially for \hcch\ are lowered
significantly.  Some HNCO lines overlap high-$J$ lines of \hcch,
which means that more hot \hcch\ is necessary to produce the
observed absorption when HNCO is not included in the model.  In our
final model including all of the observed molecules, our derived
column densities of \hcch, $N = 5.8 \times 10^{16}$ \cmsq, and HCN,
$N = 6.9 \times 10^{16}$ \cmsq, (see Table \ref{tabparms}) are
larger than those derived in \citeauthor{lahuis00}, $N_{\rm C_2H_2}
= 8 \times 10^{15}$ \cmsq\ and $N_{\rm HCN} = 1.0 \times 10^{16}$
\cmsq. Part of the difference is due to the different Doppler
parameter used: $b = 5$~\kms\ is used in \citeauthor{lahuis00},
whereas, we derive $b \sim 1$~\kms.  We derive lower temperatures,
$T_{\rm C_2H_2} = 190-230$ K and $T_{\rm HCN} = 250-450$ K than
derived by \citeauthor{lahuis00}, $T_{\rm C_2H_2} = 800$ K and
$T_{\rm HCN} = 600$ K, and \citet{boonman03}, $T_{\rm C_2H_2} = 500$
K. We also find that we need a small covering factor to account for
the saturated yet shallow lines. The level of saturation is
difficult to determine from the \iso\ data. In order to verify that
our fit to the TEXES data is also consistent with the \iso\
observations, we take our model at the \iso-SWS resolution and
compare to the data (Figure \ref{fig:isofit}).  The model matches
the fundamental Q-branches for the two molecules as well as a hot
band Q-branch of \hcch.

\hcch\ is expected to be in rotational local thermal equilibrium
(LTE) since it has no permanent dipole moment. The fact that the two
excited vibrational states of \hcch\ from which absorption was
observed, $\nu_4$ and $\nu_5$, are found to be populated near LTE at
the \hcch\ rotational temperature, or even slightly higher, provides
information about the physical conditions of the gas.  For both
bands, the hot bands indicate they are slightly out of LTE since the
vibrational temperatures ($T_v$) are higher than the rotational
temperatures ($T_r$). For the \nuff\ -- \nufo\ bands, the $T_v$
temperatures for the two components were 346 K and 271 K while for
the 2\nufi~--~\nufi\ bands, they were 277 K and 227 K. The rotational
temperatures for the \nufi\ fundamental are 225 K and 191 K for the
two components. In LTE, the vibrational temperatures for the hot
bands would equal the rotational temperature measured by the
fundamental band lines.

The $\nu_4$ level has no allowed radiative transitions to the ground
vibrational level, and none of the rotational levels of the ground
vibrational level are coupled radiatively, in both cases due to the
symmetry of the \hcch\ molecule. This suggests that the populations
of the ground and \nufo\ states should be set by collisions, and so
should be in LTE at the same temperature. However, the \nufo\ level
can be radiatively pumped by absorption in the 7.6~\um\ \nuff\ band
followed by emission of photons in the \nuff~--~\nufo\ band. The
$\nu_5$ state must be populated predominantly radiatively for any
plausible density, since the critical density is $> 10^{10}$ \cmc.
From these considerations, we conclude that the brightness
temperature of the radiation field seen by the observed gas must be
close to the kinetic temperature of the gas. In addition, the fact
that the measured rotational temperature for HCN is similar to that
of \hcch\ indicates that the rotational levels of HCN are kept
populated either by collisions, which would require a gas density
$>~10^{7}$~\cmc, or by infrared pumping through the $\nu_2$
transitions. \citet{lahuis00} conclude that either excitation mode
is possible. The presence of \hcch\ 2\nufi~--~\nufi\ absorption
requires infrared pumping, although \nuff~--~\nufo\ absorption does
not.

We can attempt to estimate the extent to which emission in our lines
fills in the absorption, at least for \hcch, by using the
2\nufi~--~\nufi\ lines as probes of the \nufi\ population. Our
spectra require a \nufi\ vibrational temperature $\sim$300~K. We
need to compare this number to the blackbody brightness temperature
of the background continuum radiation. From the shape of the
8--13~\um\ spectrum, \citeauthor{willner76} derived a dust
temperature of 330~K assuming the emitting dust is optically thick,
or 370~K assuming optically thin silicate emission. He preferred the
latter model, which gives a brightness temperature only when
combined with an assumption about the solid angle subtended by the
source. In either case, our vibrational temperature is only
$\sim$10\% less than the background temperature, so that the source
function is $\sim$30\% below the background intensity, and lines
would be expected to saturate at $\sim$70\% of the continuum.
However, the continuum spectrum from which \citeauthor{willner76}
derived a dust temperature is the average over the continuum source.
If the gas we observe covers only a small fraction of the source,
the brightness temperature of the radiation passing through that gas
could be much larger than the average.

Given the many assumptions and uncertainties in this calculation, it
is difficult to make a strong statement.  The level at which our
\hcch\ lines saturate, $\sim$85\% of the continuum, which we model
with a small covering factor, could instead be due partly or
entirely to emission by the absorbing gas.  It is possible for the
lines to saturate at 85\% of the continuum even with small covering
factors (e.g., 24\% and 6\% for \hcch) since the line widths are
narrower than our spectral resolution. Presumably the same statement
applies to the other observed molecules.

\subsection{\nhhh}

From our infrared observations, we find that \nhhh\ has two
components: one at --57.3 \kms\ and the other at --60.1 \kms.  Most
of the \nhhh\ is found in the first component, for which the
temperature is $\sim$ 278 K and the column density is 5.2 $\times
10^{16}$ \cmsq.  The second component has a lower temperature
($\sim$248 K) and the column density is 2.8 $\times 10^{16}$ \cmsq
(see Table \ref{tabparms}). \nhhh\ absorption toward \irs\ has also
been studied in the radio by \citet{wilson83} and \citet*{henkel84}.
They found $N_{\rm NH_3} = 2 \times 10^{18}$~\cmsq\ in gas with $T$
= 170--220 K and $V_{\rm LSR} = -60$~\kms. Although their
temperature and Doppler shift is in reasonable agreement with our
bluer absorption component, their column density is nearly 20 times
ours. Without constraining the $^{15}$N/$^{14}$N isotopic ratio, our
data allow an \nhhh\ column as large as that from radio
observations, but with this \nhhh\ column $^{15}$\nhhh\ lines should
have been apparent in our spectra. It is possible that we sample
only a fraction of the column observed at cm wavelengths as a result
of larger dust opacity at \mir\ wavelengths preventing us from
observing lines of sight through the densest regions of a knotty gas
distribution. It may be notable that \citeauthor{wilson83} found, as
did we, that although relative depths of lines (in their case
hyperfine components) require large optical depths, line shapes do
not appear flat-bottomed as would be expected if they are optically
thick. They suggest that the absorbing gas may be in small
spectrally and spatially unresolved knots, each of which is
optically thick, and which combine to make a line that does not have
the shape of a thick line. Our use of two components in our fit may
be a way of approximating this situation.

\subsection{HNCO}

The velocities for the two components are --57.2 and --60.2 \kms,
with corresponding temperatures of 319 and 171 K and column densities
of  4$\times 10^{15}$ and 1$\times 10^{15}$ \cmsq.  Since the HNCO
temperatures and velocities derived from our observations agree with
the values for the other molecules, we can assume that HNCO is in
the same gas.  It appears the region probed by HNCO is very small in
angular size. \citet{zinchenko00} detect HNCO emission in the radio
and derive a beam-averaged column density of 1.1$\times$10$^{14}$
\cmsq. This value is an order of magnitude lower than our observed
column density. However, beam dilution in the radio observations can
explain the difference in the column densities.

\subsection{\chhh}

\chhh\ is detected for the first time toward warm, dense gas.  The
parameters derived for this molecule do not agree as well with the
other molecules, although the weakness of the observed lines makes
the derived parameters rather uncertain. The separation between the
centroid velocities for the two components, $V_{LSR}$ = -54.2 and
--62.8, is nearly twice the separation found for the other
molecules. Also, the covering factor for both of the components is
consistent with 100\%, whereas as it is closer to 10\% for the other
molecules. In addition, the temperature of the second component is
much hotter ($>$900 K) than the temperatures derived for the other
molecules. We note though that the fact that the observed lines are
of high excitation makes them particularly sensitive to a hot gas
component, which may not have been noticed for other molecules. And
the weakness of the lines makes the derived covering factor (which
is inferred from relative line depths indicating saturation) very
uncertain.  

\subsection{\chhhh\ and CS}
\label{sec:ch4cs}

\chhhh\ gas-phase absorption had been observed toward the
neighboring protostar IRS 9 \citep{lacy91, boo04meth} but had not
previously been seen toward \irs. We find centroid velocities of
--56.3 and --60.1 \kms, which are similar to the other molecules.
The temperatures are higher than for most molecules ($\sim$ 670 K
for both components) but have large uncertainties (see Table
\ref{tabparms}).  The best fit seems to indicate that one of the
components covers the entire source whereas the other covers about
6\% of the source. Of all of the observed molecules, \chhhh\ has the
highest column density.  The implications of the high \chhhh\ column density
the chemistry are discussed in \S \ref{sec:chemmod}.

We detect only 6 lines of CS (see Table \ref{linelist} -- full table
available online only). The lines are rather weak and parameters are
poorly constrained. Despite uncertainties, the parameters agree with
those for other molecules. The centroid velocities are at --55.2 and
--59.4 \kms. The temperatures, $T_1 = 224~K$ and $T_2 = 249~K$, are
similar to those found other molecules (except for \chhhh\ and
\chhh). From radio studies of the envelope around \irs, a CS
abundance of $\sim 10^{-10}$ is found \citep{plume97, mueller02,
shirley03}. However, also using radio observations, \citet{vdt00}
derive a CS abundance of $\sim 10^{-8}$.  Our abundance is about 2
orders of magnitude larger than the largest estimate of the
abundance in the envelope indicating that our observations are
probing material closer to the protostar.

\section{Models}
\label{sec:mod}

\subsection{Chemical Models}
\label{sec:chemmod}


We now consider the implications of the observations for the
chemistry of the observed material.  Hot core models can be used to
represent two of the scenarios discussed in \S \ref{sec:intro}: 1) material in
a circumstellar disk and 2) photo-evaporating knots of neutral
molecular material.  Hot cores are small ($r<0.1$ pc), dense
($n_{\rm H_2} > 10^7$ \cmc) and hot ($T > 100$ K) regions associated
with high-mass young stellar objects \citep{kurtz00}.  Chemically, hot cores
are identified by an enhanced abundance of fully hydrogenated
molecules such as \hho\ and \nhhh, which are usually observed
through rotational transitions at sub-millimeter and millimeter
wavelengths.  These abundances are enhanced with respect to the cold
molecular envelope. A hot core is thought to form as a result of the
protostar heating nearby material, which triggers ice sublimation
from the grain mantles \citep{charnley92}.  HNCO is another hot core
molecule which has been found in the Orion hot core with an
abundance $\sim 10^{-8}$ relative to \hh\ \citep{zinchenko00}.  

Gas-phase HNCO is possibly coming from
evaporation of grain mantles. The 4.62 \micron\ feature observed
toward some protostars has been attributed to \ocn\ frozen in grain
mantles \citep[e.g.,][]{pendleton99, vb04}. Recent \iso\ results of
solid state features toward protostars show an \ocn\ upper limit of
10$^{16}$ \cmsq\ toward IRS~1 \citep{gibb04}. The total HNCO column
derived in this work is 5.4$\times 10^{15}$ \cmsq, a factor of 2
lower than the solid \ocn\ limit.  In general, IRS~1 does not show
many ice features compared to the colder nearby source IRS~9
\citep[see][]{gibb04}, which has a column density $N$(\ocn) =
1.2$\times 10^{17}$ \cmsq.  If solid \ocn\ sublimates and captures a
proton to form HNCO, the expected column density for HNCO would be
comparable to the observed \ocn\ ice toward cold lines of sight
(like IRS 9).  However, the gas-phase column observed is at least an
order of magnitude smaller than that of the solid state ion toward
sources like IRS 9.  This indicates that when \ocn\ sublimates, a
fraction of the ices goes into forming HNCO, while most of HNCO is
destroyed in the warm gas-phase chemistry on short time scales.
Further studies of how \ocn\ reacts in the gas-phase are needed in
order to understand the difference between ice and gas phase column
densities.

Similarly, \chhhh\ is probably also 
evaporating from the grain mantles, and thus the
gas-phase abundance is highly enhanced compared with the cold
molecular envelope. Studies of the solid \chhhh\ content toward
\irs\ show that the column density is 14 times less than our
observed gas-phase column \citep{gibb04}, suggesting that the solid
\chhhh\ on grains has sublimated.  In comparison, IRS 9 has a higher 
content of solid \chhhh\ than gas-phase \chhhh\ \citep{boo04meth, gibb04}. 
If we consider that \irs\ is likely a more evolved object than IRS 9 
\citep{elmegreen77}, the higher gas-phase \chhhh\ toward \irs\ also
points to grain mantle evaporation.  The high abundance of \chhhh\ 
means that daughter products such as \chhh\ and \chh\ should be abundant. 
While the photo-dissociation of \chhhh\ preferentially forms \chh\  rather than \chhh,  
our observations of \chhh\ agree with this branching ratio.  We 
predict that \chh\ is also present toward \irs\ with a higher abundance 
than \chhh.  It is uncertain where \hcch\ is formed in the gas-phase 
chemistry or is evaporating from grain mantles.  However, most models 
do not predict the observed column densities from formation in gas-phase 
chemistry only.

\begin{figure}
 \begin{center}
 \epsfig{file=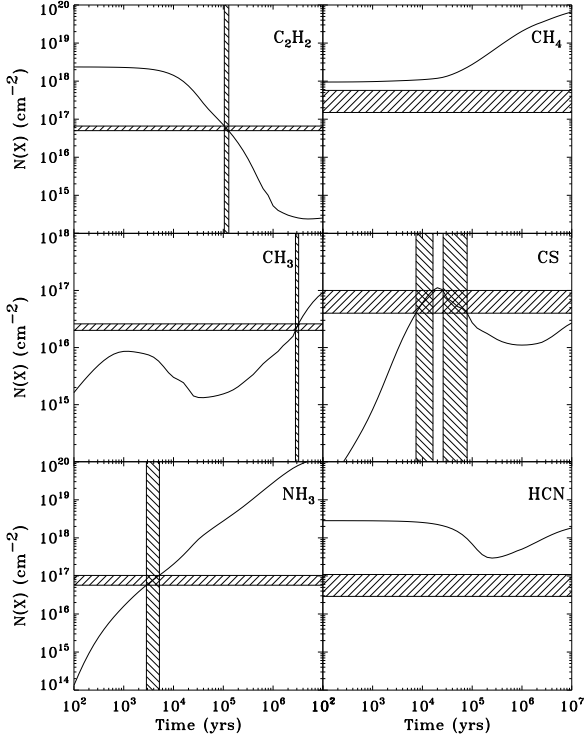, angle= 0,width=2.7in}
 \caption[Comparison to chemical models]
{These plots illustrate the determination of ages based on the column
densities of the various molecules compared with time dependent chemical model. 
The model data are from \citet{nomura04}.   The horizontal lines indicate the range in 
column densities derived in this paper.  The vertical lines indicate the age range corresponding 
to those column densities.  Note that for \chhhh\ and HCN the model overpredicts the column 
density at all ages.  
 \label{fig:chemmod}}
 \end{center}
\end{figure}

From our observations we see a high abundance of the parent
molecules \nhhh\ and \chhhh. High abundances of the other molecules
are also observed. \citet{nomura04} present a time dependent
chemical evolution of molecules in a hot core.  The models begin at
the time when the protostar turns on ($t = 0$ yrs).  Table
\ref{tab:mods} shows the abundances with respect to CO at an age of
10$^4$ years. The hot core model seems to underpredict abundances
with respect to CO of all our observed molecules. However, as
mentioned in \S \ref{sec:c2h2}, the CO column density found by
\citet{mitchell90} may not be the appropriate number with which to
compare our observations.  If the CO column along the lines of sight
containing our molecules is larger than the column measured by
\citeauthor{mitchell90}, then perhaps the abundances from the model
will more closely resemble the observations.  We can also compare
the relative abundances between our observed molecules, avoiding the
uncertainties in whether the CO and dust absorption trace the lines
of sight that our molecules are found in.  The observed and
predicted values for $N$(\chhh)/$N$(\chhhh) are similar, $\sim$0.01.
For the other molecules the ratios from the models at an age of
10$^4$ years are very different from the observed values.   The values 
for the models assume that we can see all the material and comparing 
the model results directly with the observations without accounting for 
line-of-sight does not make for a good comparison.  Figure \ref{fig:chemmod} 
shows the column density evolution for 6 of the detected molecules as determined
by \citet{nomura04} corrected for the observed beam \citetext{priv. comm.}.  For each molecule, 
the observed range of column densities is shown in the horizontal shaded area. 
This hot core model can predict the observed column densities for 4 of the 6 molecules
plotted.    However, the ages indicated by each molecule do not give a consistent 
overall age.  For \chhhh\ and HCN the model overpredicts the observed column densities.  
The other molecules suggest ages between 2$\times 10^{3}$ and 2$\times 10^{6}$ years. 
While ages vary by three orders of magnitude, it should be noted that the 
model has not been optimized to be an accurate representation of 
this source.  With chemical models that more closely represent 
the source, it may be possible to determine a more consistent age.

\begin{deluxetable}{lccc}
\tablecolumns{4}
\tablecaption{{$N$(X)/$N$(CO) from chemical models}}
 \tablewidth{0pt}
 \tablehead{
  \colhead{Molecule}            &
  \colhead{Observed}      &
  \colhead{Hot Core Model$^a$} &
  \colhead{Disk Models$^b$}            \\
  \colhead{}                    &
  \colhead{(10$^{-3}$)}              &
  \colhead{(10$^{-3}$)}     &
  \colhead{(10$^{-3}$)}
} \startdata
\hcch  & 5.7 & 0.88   & 0.06 -- 0.1    \\
HCN     & 6.9 & 0.02  & 16 -- 24   \\
\chhh      & 2.3  & 0.016  &  0.02 -- 1.6   \\
\nhhh  &  8.1 & 0.76   & 1 -- 6  \\
\chhhh & 36  & 1.6  &  7 -- 35 \\
CS$^*$    & 7.0  & 0.0031 & 0.03 -- 0.07 \\
 \label{tab:mods}
\enddata

\tablecomments{The model values were taken at 10$^4$ years.}

 \tablerefs{$^a$ \citealt{nomura04} and H. Nomura, private
communication 2004 \\ $^b$ \citealt{nguyen02}}

\end{deluxetable}

%

If we consider the disk scenario, we can also compare our
observations to a chemical model of a photo-evaporating disk.
\citet{nguyen02} present a chemical model for a disk around a 10
M$_{\odot}$ star. The model consists of a flared disk with a
photo-evaporating layer on the surface.  Table \ref{tab:mods}
compares our derived abundances with respect to CO for the various
molecules to the predicted values from disk models. The summarized
results of models taking the midplane temperature for the
temperature of the layers below the photo-dissociation region (PDR)
are discussed here. The difference among the three models is the
variation in the X-ray ionization rate, $\zeta$ (1$\zeta$ =
1.3$\times$10$^{-17}$s$^{-1}$), from 1 to 10$^5 \zeta$.  Table
\ref{tab:mods} shows that the abundance for \hcch\ is
underpredicted. In the chemical model, \hcch\ is produced only by
gas phase reactions. One possible explanation is that \hcch\ is
formed as ice on grain mantles.  The abundance of \nhhh\ is also
underpredicted. The model with the highest X-ray ionization rate
produces \chhh\ abundances close to the observed value.  For that
model, $N$(\chhh)/$N$(\chhhh) is 4 times larger than the observed
ratio.  These models do not fit our observations very
well but can serve as a guide to the expected abundances in such a
scenario. The model by \citeauthor{nguyen02} has not been modified
to try to fit data. So, adapting the model for this object may give
predictions closer to the observations. The star in NGC 7538 IRS 1
is believed to be a 30 M$_{\odot}$ star. The different radiation
field may affect the resulting chemistry.

\citet{doty02} present a model of chemical evolution of the
envelopes of massive protostars using AFGL 2591 as an example.  From
that model, they conclude that enhancements of \hcch, HCN and
\chhhh\ are possible at late times in the evolution ($> 10^{5}$
years) and at high temperature ($T \sim$ 800 K).  Our derived
temperatures indicate cooler material, yet the molecules seem to be
enhanced in the line of sight. Alternatively, these species could be
produced at high temperature in the inner disk, and subsequently
brought outward by the disk wind.  Including UV and X-ray radiation
in the model affects the chemistry especially relating to HCN.
\cite{stauber04, stauber05} find that increasing the UV and X-ray
flux can lead to enhanced HCN starting by the ionization of N$_2$.
The column densities for HCN and CS from the \citeauthor{stauber05}
model of protostar AFGL 2591 agree with our observed abundances
while their column densities for \hcch, \chhhh, and \nhhh\ are lower
than our observed values.

Hot core and disk chemistry models predict the enhancement of
molecules such as \hcch, \chhhh, and \nhhh.  However, the observed
abundance of \hcch\ is higher than what models predict from warm
gas-phase chemistry (see Table \ref{tab:mods}).  This indicates that
\hcch\ is probably frozen on dust grains and sublimates along with
molecules like \chhhh.  \citet{boudin98} studied the solid features
of \hcch, especially when mixed with \hho\ or CO.  They found that
the \hcch\ features broaden substantially when mixed with \hho\ and
to a lesser extent when mixed with CO.  They compare the laboratory
data to \iso\ data for the colder neighbor, IRS~9, which resulted in
an upper limit of $8 \times 10^{17}$ \cmsq\ for the column density
of solid \hcch.  This upper limit is consistent with the observed
gas-phase column density seen toward \irs.  So, it is possible that
solid \hcch\ is sublimating from grain mantles as protostars heat
the environment.  The observed \chhh/\chhhh\ ratio agrees with the
branching ratio for the destruction of \chhhh. Based on these
results \chh\ should also be very abundant.

\subsection{Physical Models}
\label{physmod}

\begin{figure}
 \begin{center}
 \epsfig{file=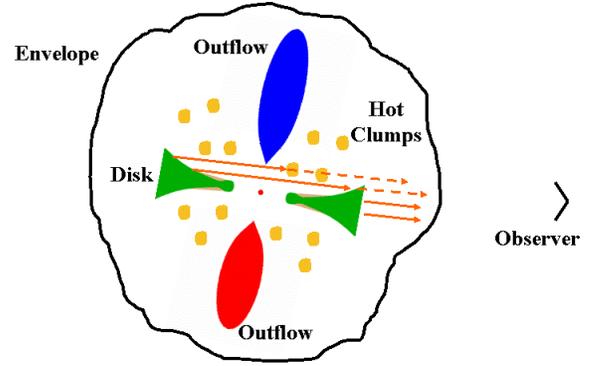, angle= 0,width=3.2in}
 \caption[Cartoon representation of the \ngc\ \irs\ region]
{A cartoon representation of the IRS 1 region. The absorbing
material containing the molecules observed can either be located in
the disk or in the hot knots, while the continuum is coming from
the photo-evaporating layers of the the inner rim and surface of the
disk.  Flaring is not to scale.  The continuum can also be from
other parts of the disk for which lines of sight do not cross the
material with our molecules. There may also be more extended dust
emission from the envelope that contributes to the continuum but
does not contain our molecules in high abundances.  In this cartoon, 
North roughly points toward the top of the page. 
 \label{fig:cartoon}}
 \end{center}
\end{figure}

We now attempt to construct a physical and geometrical model of
\ngc\ \irs.  Within the possible scenarios presented from various
radio and infrared observations \citep[e.g.,][]{minier01, lugo04,
debuizer05, kraus06}, we propose a scenario in which the molecular
absorption presented here comes from a circumstellar disk.  We will
examine other possibilities first.  Figure \ref{fig:cartoon} depicts
the possible scenarios allowed by the available observations.

We can quickly rule out a simple model in which the absorbing
molecular gas is in the foreground molecular cloud and not closely
associated with \irs. Our observations require the gas to be much
hotter ($T \sim$ 300~K) and denser ($n_{\rm H_2} \sim 10^7$~\cmc\ to
maintain rotational LTE of HCN out to $J = 21$) than is found away
from luminous sources in molecular clouds.  According to \citet{vdt00}, 
temperatures do not reach the observed values until you get to within 
400 AU of the central star.  
In addition, the \ir\
radiation field must have a brightness temperature $\sim$300~K to
populate the \hcch\ $\nu_5$ level sufficiently to account for the
observed 2\nufi --\nufi\ absorption. Undoubtedly,
the absorbing molecular gas is in close proximity to the \ir\
continuum source \irs.

It is not quite so easy to rule out a model in which the absorbing
molecules are in boundary region between the  envelope around the \irs\
hypercompact H~II region  and the outflow. This gas could be compressed by the
ionized wind, and if it is as close as 0.1\arcsec, or 280~AU, from a
$10^{5}~L_{\odot}$ source it would have a temperature near 300~K.
The temperature structure determined by \citet{vdt00} indicates that
indeed temperatures range from 200 -- 400 K at radii of 220--380 AU.
However, interaction with the ionized wind, which has a velocity
$\sim$100~\kms, would be expected to accelerate the gas, causing
broad, blue-shifted absorption  \citep[see][]{vdt00}. In contrast, the observed lines have
widths of $<8$~\kms\ and have centroids within a few \kms\ of those
seen in surrounding molecular gas. Another argument against the
presence of the observed gas in an envelope around \irs\ is the fact
that net absorption is seen. Since the vibrational temperature is
comparable to the brightness temperature of the continuum radiation,
emission lines would be seen if the molecular gas had a larger
extent than the continuum source. If the lines arose in molecular
shell surrounding the hypercompact H~II region, this probably would
be the case.

On the other hand, \citet{campbell84} proposes that the centimeter
continuum from IRS 1 results from partially ionized material in an
outflow. The centimeter continuum emission has been spatially
resolved into knots by \citet{gaume95}, who suggest that the
emission comes from photo-evaporation of knots of neutral molecular
material (see Fig. \ref{fig:cartoon}). The \methanol\ masers seen by
\citet{minier00} also seem to trace the knotty structure (in addition to 
the disk described in \S \ref{sec:intro}) probed by
the cm continuum emission. The knots may be material stripped from
the disk.  The stellar wind would then blow on these knots. However,
if our molecules are in these knots
we would expect large velocity difference between the knots
resulting in broad lines, which are not observed. If we compare the
velocities we derive for the two components of our molecules to
those observed for the \methanol\ masers, we find that they match
the two velocities found for the A cluster of masers. This indicates
that the infrared observations may be probing the same gas as the
\methanol\ masers.   This scenario is further supported by \citet{kraus06}.
However, \citeauthor{kraus06} also find that the interpretation of the 
masers tracing a disk also agrees with their observations. 

We propose that the absorbing material is in a circumstellar disk.
There is evidence for a disk close to edge-on from other
observations and models \citep[e.g.,][]{pestalozzi04, lugo04,
debuizer05}. In this situation, hot dust in the surface layers or
the inner rim of a flared disk could provide the continuum source.
With a nearly edge-on orientation, radiation from the inner region
could pass through the outer disk atmosphere, where absorption lines
could be formed.  Emission from some of the dust on the surface of
a slightly inclined disk or in an outflow could be observed
directly, accounting for the dilution of the spectrum that we
modeled with a small covering factor.  This scenario is similar to
the one described by \cite{lahuis06} to describe \hcch\ and HCN
absorption toward the low-mass protostar IRS 46 in Ophiuchus (YLW 16B).  
Detailed modeling
of the disk structure of IRS 46 showed that it was possible to have
the absorbing material in the disk illuminated by the continuum from
the surface.  Figure \ref{fig:cartoon} is a cartoon of our proposed
picture, where the orange rays represent the continuum. Some of the
continuum radiation passes through the absorbing outer disk.  In
this scenario, dust settling is likely to occur, allowing for
observations of large columns of gas without too much dust.

\section{Conclusions}
\label{sec:conc}

\noindent Here we present a summary of our conclusions:

1. \ngc~\irs\ has a rich spectrum in the \mir.  At high spectral
resolution, lines of previously observed molecules such as \hcch\
and HCN are seen as well as weak lines of molecules such as HNCO,
\chhh, \nhhh, \chhhh, and CS.  We have presented the first infrared
detection of interstellar HNCO.  This is also the first detection of
\chhh\ toward dense gas.

2. The data show shallow yet saturated lines indicating that either
the absorbing gas does not fully cover the continuum source or there
is emission filling in the absorption.  The observed ortho:para
equivalent width ratio for \hcch\ is $\sim$ 1.5 instead of the
expected value of 3.

3. The rotational temperatures for the various molecules range
between 200 and 400 K with the exception of \chhh, which has one hot
component ($T \sim 900$ K). From the derived temperatures, we find
that the gas is within $\sim$ 400 AU of the central star
\citep{vdt00}.

4. We suggest that the material traces a close to edge-on
circumstellar disk as determined from other observations
\citep[e.g.,][]{debuizer05}. However, it can also be tracing hot,
knotty material close in to the star (see Figure \ref{fig:cartoon}).

5. Because it is possible to use hot core chemistry to describe
chemistry in a disk, the different chemical models do not help
identify the physical location of the absorbing gas.  However, the
models do seem to indicate that \ngc\ \irs\ is an evolved protostar
with an age $\sim 10^{5}$ years.

6. Our observations help constrain the chemistry in massive
protostars. The abundances for all the molecules are enhanced,
except for possibly CS (see discussion in \S~ \ref{sec:ch4cs}).
Species known to be present in icy mantles such as \chhhh\ and
\nhhh\ have high column densities suggesting that they have recently
sublimated from grain mantles. The high abundance of the daughter
molecules such \chhh\ gives constraints on the gas-phase chemistry
happening after sublimation of ices. HNCO also provides some
constraints about the presence of \ocn\ on dust grains.  It is
unclear whether \hcch\ is a parent molecule or a product of warm
gas-phase chemistry.  For molecules like HCN, UV/X-ray radiation is
important in producing the observed abundances.

\acknowledgements{CK would like to thank A.C.A. Boogert, A. M. S.
Boonman-Gloudemans, S. Doty and F. Lahuis for useful discussions.
The authors would like to thank T. K. Greathouse for assisting with
the observations. CK acknowledges support from the NASA Astrobiology 
Program under RTOP 344-53-51. This work was partly supported by NSF grant
AST-0607312. }

\begin{appendix}
\label{app}
\section{Spectroscopy of \chhh\ and HNCO}

\chhh\ is a nearly planar molecule and thus do not expect to see
inversion transitions as observed in \nhhh.  We observe the
out-of-plane vibrational mode (similar to the umbrella mode in
\nhhh).  We observe splitting due to spin-rotation interactions. For
symmetric-top molecules, the splitting of a rotational level is
given by

\begin{equation}
\Delta \nu = (N+1/2)\{ \epsilon^{^{(\nu)}}\!\!\!\!\!_{_{bb}} -
(\epsilon^{^{ (\nu)}}\!\!\!\!\!_{_{bb}} -
\epsilon^{^{(\nu)}}\!\!\!\!\!_{_{cc}})K^2/[ N(N+1)]\},
 \label{eq:ch3}
\end{equation}

\noindent where $\epsilon^{^{(\nu)}}\!\!\!\!\!_{_{bb}}$ and
$\epsilon^{^{ (\nu)}}\!\!\!\!\!_{_{cc}}$ are the spin-rotation
coupling constants for the B and C axes, respectively, at a given
vibrational state, $\nu$.  We use Equation (\ref{eq:ch3}) in
deriving the splitting for the lines we observed. The values for
$\epsilon^{^{(\nu)}}\!\!\!\!\!_{_{bb}}$ and $\epsilon^{
^{(\nu)}}\!\!\!\!\!_{_{cc}}$ were taken from \citet{yamada81}.

HNCO is a quasilinear, nearly symmetric-top molecule. The moment of
inertia about the figure axis is small and the moments of inertia
about the other two axes are much larger ($\sim$ 100 times larger)
and about equal to each other. Because it is a slightly asymmetric
molecule, the levels with K $\ne$ 0 are split into two components.
\citet{steiner79} refer to these levels as upper (U) and lower (L).  Table \ref{rules}  shows the selection 
rules for HNCO.


\begin{table}
\begin{center}
\caption{HNCO Selection Rules \label{rules}}
\begin{tabular}{llcll}
\tableline\tableline
  \multicolumn{2}{c}{$\Delta K = 0$} &    &
\multicolumn{2}{c}{$\Delta K = \pm 1$} \\
\cline{1-2} \cline{4-5} 
 {$ K \ne 0$} &  &  &  & \\
 $\Delta J = 0$   & {$\Delta J = \pm 1$}    &  &
 {$\Delta J = 0$}    &  {$\Delta J  = \pm 1$} \\
 $U \leftrightarrow L$ & $U \leftrightarrow U$ & & $U \leftrightarrow U$ & $U \leftrightarrow L$  \\
  & $L \leftrightarrow L$  & & $L \leftrightarrow L$  & \\
\tableline
\end{tabular}
\end{center}
\end{table}


\noindent In Figure \ref{fig:hncoblowup}, the HNCO lines labeled
R(J)1U correspond to lines where $\Delta K = 0$ and $\Delta J = 1$
and K = 1 upper. Likewise, the lines labeled R(J)1L are lines where
$\Delta K = 0$ and $\Delta J = 1$ and K = 1 lower. Because the
splitting between the two levels is small the lines overlap in the
spectral region shown in Figures \ref{fig:hncoall} and
\ref{fig:hncoblowup}. Some lines from the upper and lower K states
are blended together and give the appearance of double peaked lines.

\end{appendix}


\begin{thebibliography} 
\expandafter\ifx\csname natexlab\endcsname\relax\def\natexlab#1{#1}\fi

\bibitem[{{Bevington} \& {Robinson}(2003)}]{bev03}
{Bevington}, P.~R., \& {Robinson}, D.~K. 2003, {Data reduction and error
  analysis for the physical sciences} (3rd ed.; Boston: McGraw-Hill)

\bibitem[{{Boogert} {et~al.}(2004){Boogert}, {Blake}, \&
  {{\"O}berg}}]{boo04meth}
{Boogert}, A.~C.~A., {Blake}, G.~A., \& {{\"O}berg}, K. 2004, \apj, 615, 344

\bibitem[{{Boonman} \& {van Dishoeck}(2003)}]{boonman03h2o}
{Boonman}, A.~M.~S., \& {van Dishoeck}, E.~F. 2003, \aap, 403, 1003

\bibitem[{{Boonman} {et~al.}(2003){Boonman}, {van Dishoeck}, {Lahuis}, \&
  {Doty}}]{boonman03}
{Boonman}, A.~M.~S., {van Dishoeck}, E.~F., {Lahuis}, F., \& {Doty}, S.~D.
  2003, \aap, 399, 1063

\bibitem[{{Botschwina} \& {Sebald}(1985)}]{bot85}
{Botschwina}, P., \& {Sebald}, P. 1985, Journal of Molecular Spectroscopy, 110,
  1

\bibitem[{{Boudin} {et~al.}(1998){Boudin}, {Schutte}, \&
  {Greenberg}}]{boudin98}
{Boudin}, N., {Schutte}, W.~A., \& {Greenberg}, J.~M. 1998, \aap, 331, 749

\bibitem[{{Campbell}(1984)}]{campbell84}
{Campbell}, B. 1984, \apjl, 282, L27

\bibitem[{{Campbell} \& {Thompson}(1984)}]{campthomp84}
{Campbell}, B., \& {Thompson}, R.~I. 1984, \apj, 279, 650

\bibitem[{{Charnley} {et~al.}(1992){Charnley}, {Tielens}, \&
  {Millar}}]{charnley92}
{Charnley}, S.~B., {Tielens}, A.~G.~G.~M., \& {Millar}, T.~J. 1992, \apjl, 399,
  L71

\bibitem[{{De Buizer} \& {Minier}(2005)}]{debuizer05}
{De Buizer}, J.~M., \& {Minier}, V. 2005, \apjl, 628, L151

\bibitem[{{Dickel} {et~al.}(1982){Dickel}, {Rots}, {Goss}, \&
  {Forster}}]{dickel82}
{Dickel}, H.~R., {Rots}, A.~H., {Goss}, W.~M., \& {Forster}, J.~R. 1982,
  \mnras, 198, 265

\bibitem[{{Doty} {et~al.}(2002){Doty}, {van Dishoeck}, {van der Tak}, \&
  {Boonman}}]{doty02}
{Doty}, S.~D., {van Dishoeck}, E.~F., {van der Tak}, F.~F.~S., \& {Boonman},
  A.~M.~S. 2002, \aap, 389, 446

\bibitem[{{Elmegreen} \& {Lada}(1977)}]{elmegreen77}
{Elmegreen}, B.~G., \& {Lada}, C.~J. 1977, \apj, 214, 725

\bibitem[{{Feuchtgruber} {et~al.}(2000){Feuchtgruber}, {Helmich}, {van
  Dishoeck}, \& {Wright}}]{feucht00}
{Feuchtgruber}, H., {Helmich}, F.~P., {van Dishoeck}, E.~F., \& {Wright}, C.~M.
  2000, \apjl, 535, L111

\bibitem[{{Gaume} {et~al.}(1995){Gaume}, {Goss}, {Dickel}, {Wilson}, \&
  {Johnston}}]{gaume95}
{Gaume}, R.~A., {Goss}, W.~M., {Dickel}, H.~R., {Wilson}, T.~L., \& {Johnston},
  K.~J. 1995, \apj, 438, 776

\bibitem[{{Gibb} {et~al.}(2004){Gibb}, {Whittet}, {Boogert}, \&
  {Tielens}}]{gibb04}
{Gibb}, E.~L., {Whittet}, D.~C.~B., {Boogert}, A.~C.~A., \& {Tielens},
  A.~G.~G.~M. 2004, \apjs, 151, 35

\bibitem[{{Henkel} {et~al.}(1984){Henkel}, {Wilson}, \& {Johnston}}]{henkel84}
{Henkel}, C., {Wilson}, T.~L., \& {Johnston}, K.~J. 1984, \apjl, 282, L93

\bibitem[{{Hoffman} {et~al.}(2003){Hoffman}, {Goss}, {Palmer}, \&
  {Richards}}]{hoffman03}
{Hoffman}, I.~M., {Goss}, W.~M., {Palmer}, P., \& {Richards}, A.~M.~S. 2003,
  \apj, 598, 1061

\bibitem[{{Kameya} {et~al.}(1990){Kameya}, {Morita}, {Kawabe}, \&
  {Ishiguro}}]{kameya90}
{Kameya}, O., {Morita}, K.-I., {Kawabe}, R., \& {Ishiguro}, M. 1990, \apj, 355,
  562

\bibitem[{{Kraus} {et~al.}(2006){Kraus}, {Balega}, {Elitzur}, {Hofmann},
  {Meyer}, {Preibisch}, {Rosen}, {Schertl}, {Weigelt}, \& {Young}}]{kraus06}
{Kraus}, S., {Balega}, Y., {Elitzur}, M., {Hofmann}, K.-H., {Meyer}, M.,
  {Preibisch}, T., {Rosen}, A., {Schertl}, D., {Weigelt}, G., \& {Young}, E.~T.
  2006, \aap, 1000, 446

\bibitem[{{Kurtz} {et~al.}(2000){Kurtz}, {Cesaroni}, {Churchwell}, {Hofner}, \&
  {Walmsley}}]{kurtz00}
{Kurtz}, S., {Cesaroni}, R., {Churchwell}, E., {Hofner}, P., \& {Walmsley},
  C.~M. 2000, Protostars and Planets IV, 299

\bibitem[{{Lacy} {et~al.}(1989){Lacy}, {Achtermann}, {Bruce}, {Lester},
  {Arens}, {Peck}, \& {Gaalema}}]{lacy89irsh}
{Lacy}, J.~H., {Achtermann}, J.~M., {Bruce}, D.~E., {Lester}, D.~F., {Arens},
  J.~F., {Peck}, M.~C., \& {Gaalema}, S.~D. 1989, \pasp, 101, 1166

\bibitem[{{Lacy} {et~al.}(1991){Lacy}, {Carr}, {Evans}, {Baas}, {Achtermann},
  \& {Arens}}]{lacy91}
{Lacy}, J.~H., {Carr}, J.~S., {Evans}, N.~J., {Baas}, F., {Achtermann}, J.~M.,
  \& {Arens}, J.~F. 1991, \apj, 376, 556

\bibitem[{{Lacy} {et~al.}(2002){Lacy}, {Richter}, {Greathouse}, {Jaffe}, \&
  {Zhu}}]{lacy02}
{Lacy}, J.~H., {Richter}, M.~J., {Greathouse}, T.~K., {Jaffe}, D.~T., \& {Zhu},
  Q. 2002, \pasp, 114, 153

\bibitem[{{Lahuis} \& {van Dishoeck}(2000)}]{lahuis00}
{Lahuis}, F., \& {van Dishoeck}, E.~F. 2000, \aap, 355, 699

\bibitem[{{Lahuis} {et~al.}(2006){Lahuis}, {van Dishoeck}, {Boogert},
  {Pontoppidan}, {Blake}, {Dullemond}, {Evans}, {Hogerheijde}, {J{\o}rgensen},
  {Kessler-Silacci}, \& {Knez}}]{lahuis06}
{Lahuis}, F., {van Dishoeck}, E.~F., {Boogert}, A.~C.~A., {Pontoppidan}, K.~M.,
  {Blake}, G.~A., {Dullemond}, C.~P., {Evans}, N.~J., {Hogerheijde}, M.~R.,
  {J{\o}rgensen}, J.~K., {Kessler-Silacci}, J.~E., \& {Knez}, C. 2006, \apjl,
  636, L145

\bibitem[{{Lowenthal} {et~al.}(2002){Lowenthal}, {Khanna}, \& {Moore}}]{low01}
{Lowenthal}, M.~S., {Khanna}, R.~K., \& {Moore}, M.~H. 2002, Spectrochimica
  Acta Part A, 58, 73

\bibitem[{{Lugo} {et~al.}(2004){Lugo}, {Lizano}, \& {Garay}}]{lugo04}
{Lugo}, J., {Lizano}, S., \& {Garay}, G. 2004, \apj, 614, 807

\bibitem[{{Madden} {et~al.}(1986){Madden}, {Irvine}, {Matthews}, {Brown}, \&
  {Godfrey}}]{madden86}
{Madden}, S.~C., {Irvine}, W.~M., {Matthews}, H.~E., {Brown}, R.~D., \&
  {Godfrey}, P.~D. 1986, \apjl, 300, L79

\bibitem[{{Martin}(1973)}]{martin73}
{Martin}, A.~H.~M. 1973, \mnras, 163, 141

\bibitem[{{Menten} {et~al.}(1986){Menten}, {Walmsley}, {Henkel}, {Wilson},
  {Snyder}, {Hollis}, \& {Lovas}}]{menten86}
{Menten}, K.~M., {Walmsley}, C.~M., {Henkel}, C., {Wilson}, T.~L., {Snyder},
  L.~E., {Hollis}, J.~M., \& {Lovas}, F.~J. 1986, \aap, 169, 271

\bibitem[{{Minier} {et~al.}(1998){Minier}, {Booth}, \& {Conway}}]{minier98}
{Minier}, V., {Booth}, R.~S., \& {Conway}, J.~E. 1998, \aap, 336, L5

\bibitem[{{Minier} {et~al.}(2000){Minier}, {Booth}, \& {Conway}}]{minier00}
---. 2000, \aap, 362, 1093

\bibitem[{{Minier} {et~al.}(2001){Minier}, {Conway}, \& {Booth}}]{minier01}
{Minier}, V., {Conway}, J.~E., \& {Booth}, R.~S. 2001, \aap, 369, 278

\bibitem[{{Mitchell} {et~al.}(1990){Mitchell}, {Maillard}, {Allen}, {Beer}, \&
  {Belcourt}}]{mitchell90}
{Mitchell}, G.~F., {Maillard}, J.-P., {Allen}, M., {Beer}, R., \& {Belcourt},
  K. 1990, \apj, 363, 554

\bibitem[{{Mueller} {et~al.}(2002){Mueller}, {Shirley}, {Evans}, \&
  {Jacobson}}]{mueller02}
{Mueller}, K.~E., {Shirley}, Y.~L., {Evans}, N.~J., \& {Jacobson}, H.~R. 2002,
  \apjs, 143, 469

\bibitem[{{Nguyen} {et~al.}(2002){Nguyen}, {Viti}, \& {Williams}}]{nguyen02}
{Nguyen}, T.~K., {Viti}, S., \& {Williams}, D.~A. 2002, \aap, 387, 1083

\bibitem[{{Nomura} \& {Millar}(2004)}]{nomura04}
{Nomura}, H., \& {Millar}, T.~J. 2004, \aap, 414, 409

\bibitem[{{Pendleton} {et~al.}(1999){Pendleton}, {Tielens}, {Tokunaga}, \&
  {Bernstein}}]{pendleton99}
{Pendleton}, Y.~J., {Tielens}, A.~G.~G.~M., {Tokunaga}, A.~T., \& {Bernstein},
  M.~P. 1999, \apj, 513, 294

\bibitem[{{Pestalozzi} {et~al.}(2004){Pestalozzi}, {Elitzur}, {Conway}, \&
  {Booth}}]{pestalozzi04}
{Pestalozzi}, M.~R., {Elitzur}, M., {Conway}, J.~E., \& {Booth}, R.~S. 2004,
  \apjl, 603, L113

\bibitem[{{Plume} {et~al.}(1997){Plume}, {Jaffe}, {Evans}, {Martin-Pintado}, \&
  {Gomez-Gonzalez}}]{plume97}
{Plume}, R., {Jaffe}, D.~T., {Evans}, N.~J., {Martin-Pintado}, J., \&
  {Gomez-Gonzalez}, J. 1997, \apj, 476, 730

\bibitem[{{Pratap} {et~al.}(1989){Pratap}, {Batrla}, \& {Snyder}}]{pratap89}
{Pratap}, P., {Batrla}, W., \& {Snyder}, L.~E. 1989, \apj, 341, 832

\bibitem[{{Scoville} {et~al.}(1986){Scoville}, {Sargent}, {Sanders},
  {Claussen}, {Masson}, {Lo}, \& {Phillips}}]{scoville86outflows}
{Scoville}, N.~Z., {Sargent}, A.~I., {Sanders}, D.~B., {Claussen}, M.~J.,
  {Masson}, C.~R., {Lo}, K.~Y., \& {Phillips}, T.~G. 1986, \apj, 303, 416

\bibitem[{{Shirley} {et~al.}(2003){Shirley}, {Evans}, {Young}, {Knez}, \&
  {Jaffe}}]{shirley03}
{Shirley}, Y.~L., {Evans}, N.~J., {Young}, K.~E., {Knez}, C., \& {Jaffe}, D.~T.
  2003, \apjs, 149, 375

\bibitem[{{St{\"a}uber} {et~al.}(2005){St{\"a}uber}, {Doty}, {van Dishoeck}, \&
  {Benz}}]{stauber05}
{St{\"a}uber}, P., {Doty}, S.~D., {van Dishoeck}, E.~F., \& {Benz}, A.~O. 2005,
  \aap, 440, 949

\bibitem[{{St{\"a}uber} {et~al.}(2004){St{\"a}uber}, {Doty}, {van Dishoeck},
  {J{\o}rgensen}, \& {Benz}}]{stauber04}
{St{\"a}uber}, P., {Doty}, S.~D., {van Dishoeck}, E.~F., {J{\o}rgensen}, J.~K.,
  \& {Benz}, A.~O. 2004, \aap, 425, 577

\bibitem[{{Steiner} {et~al.}(1979){Steiner}, {Wishah}, {Polo}, \&
  {McCubbin}}]{steiner79}
{Steiner}, D.~A., {Wishah}, K.~A., {Polo}, S.~R., \& {McCubbin}, T.~K. 1979, J.
  of Mol. Spec., 76, 341

\bibitem[{{Ungerechts} {et~al.}(2000){Ungerechts}, {Umbanhowar}, \&
  {Thaddeus}}]{ung00}
{Ungerechts}, H., {Umbanhowar}, P., \& {Thaddeus}, P. 2000, \apj, 537, 221

\bibitem[{{van Broekhuizen} {et~al.}(2004){van Broekhuizen}, {Keane}, \&
  {Schutte}}]{vb04}
{van Broekhuizen}, F.~A., {Keane}, J.~V., \& {Schutte}, W.~A. 2004, \aap, 415,
  425

\bibitem[{{van der Tak} {et~al.}(2000){van der Tak}, {van Dishoeck}, {Evans},
  \& {Blake}}]{vdt00}
{van der Tak}, F.~F.~S., {van Dishoeck}, E.~F., {Evans}, N.~J., \& {Blake},
  G.~A. 2000, \apj, 537, 283

\bibitem[{{Werner} {et~al.}(1979){Werner}, {Becklin}, {Gatley}, {Matthews},
  {Neugebauer}, \& {Wynn-Williams}}]{werner79}
{Werner}, M.~W., {Becklin}, E.~E., {Gatley}, I., {Matthews}, K., {Neugebauer},
  G., \& {Wynn-Williams}, C.~G. 1979, \mnras, 188, 463

\bibitem[{{Willner}(1976)}]{willner76}
{Willner}, S.~P. 1976, \apj, 206, 728

\bibitem[{{Willner} {et~al.}(1982){Willner}, {Gillett}, {Herter}, {Jones},
  {Krassner}, {Merrill}, {Pipher}, {Puetter}, {Rudy}, {Russell}, \&
  {Soifer}}]{willner82}
{Willner}, S.~P., {Gillett}, F.~C., {Herter}, T.~L., {Jones}, B., {Krassner},
  J., {Merrill}, K.~M., {Pipher}, J.~L., {Puetter}, R.~C., {Rudy}, R.~J.,
  {Russell}, R.~W., \& {Soifer}, B.~T. 1982, \apj, 253, 174

\bibitem[{{Wilson} {et~al.}(1983){Wilson}, {Walmsley}, {Batrla}, \&
  {Mauersberger}}]{wilson83}
{Wilson}, T.~L., {Walmsley}, C.~M., {Batrla}, W., \& {Mauersberger}, R. 1983,
  \aap, 127, L19

\bibitem[{{Wormhoudt} \& {McCurdy}(1989)}]{wor89}
{Wormhoudt}, J., \& {McCurdy}, K.~E. 1989, Chem. Phys. Letters, 156, 47

\bibitem[{{Wynn-Williams} {et~al.}(1974){Wynn-Williams}, {Becklin}, \&
  {Neugebauer}}]{wynnwilliams74}
{Wynn-Williams}, C.~G., {Becklin}, E.~E., \& {Neugebauer}, G. 1974, \apj, 187,
  473

\bibitem[{{Yamada} {et~al.}(1981){Yamada}, {Hirota}, \& {Kawaguchi}}]{yamada81}
{Yamada}, C., {Hirota}, E., \& {Kawaguchi}, K. 1981, \jcp, 75, 5256

\bibitem[{{Zheng} {et~al.}(2001){Zheng}, {Zhang}, {Ho}, \& {Pratap}}]{zheng01}
{Zheng}, X.-W., {Zhang}, Q., {Ho}, P.~T.~P., \& {Pratap}, P. 2001, \apj, 550,
  301

\bibitem[{{Zinchenko} {et~al.}(2000){Zinchenko}, {Henkel}, \&
  {Mao}}]{zinchenko00}
{Zinchenko}, I., {Henkel}, C., \& {Mao}, R.~Q. 2000, \aap, 361, 1079

\end{thebibliography}

\end{document}